\documentclass[aps,prb,reprint,
  showpacs,pdflatex]{revtex4-1}
\usepackage[dvipdfmx]{graphicx}
\usepackage{bm,braket}
\usepackage{color}
\usepackage{amsmath,amssymb}
\usepackage{multirow}
\usepackage{ulem}
\usepackage{siunitx}

\renewcommand{\emph}[1]{{\it #1}}

\newcommand{\tr}{\mathrm{tr}\,}
\newcommand{\sgn}{\mathrm{sgn}\,}

\newcommand{\muB}{\mu_{\rm{B}}}

\newcommand{\Eq}[1]{Eq.~\eqref{#1}}
\newcommand{\Eqs}[2]{Eqs.~\eqref{#1}~and~\eqref{#2}}
\newcommand{\Eqss}[1]{Eqs.~\eqref{#1}}
\newcommand{\Ref}[1]{Ref.~\onlinecite{#1}}
\newcommand{\Refs}[2]{Refs.~\onlinecite{#1}~and~\onlinecite{#2}}

\newcommand{\Fig}[1]{Fig.~\ref{#1}}

\newcommand{\Figss}[1]{Figs.~\ref{#1}}
\graphicspath{./figure}
\newcommand{\Sec}[1]{Sec.~\ref{#1}}
\newcommand{\Secs}[2]{Secs.~\ref{#1}~and~\ref{#2}}

\begin{document}

\title{Majorana~flat~bands, chiral~Majorana~edge~states, and~unidirectional~Majorana~edge~states in noncentrosymmetric superconductors
  }
\author{Akito Daido}
\email[]{daido@scphys.kyoto-u.ac.jp}
\author{Youichi Yanase}
\affiliation{Department of Physics, Graduate School of Science, Kyoto University, Kyoto 606-8502, Japan}

\date{\today}

\begin{abstract}
 We study two-dimensional noncentrosymmetric nodal superconductors under Zeeman field and 
clarify the field angle dependence of topological properties. 
It has been shown that the nodal excitation acquires an excitation gap 
due to the Zeeman field perpendicular to anti-symmetric spin-orbit coupling, and then 
gapful topological superconductivity is realized\cite{Daido2016}. 
We show that the system undergoes gapful-gapless transition against tilting of the field. 
The gapless phase remains to show a finite band gap and unusual Majorana edge states 
in between the bulk bands. The Majorana edge states naturally propagate in a same direction between oppositely-oriented edges. We elucidate relations of such unidirectional Majorana edge states 
with chiral edge states in the gapful topological superconducting phase and previously studied 
Majorana flat bands at zero Zeeman field. A compact formula of topological invariants characterizing 
the edge states is given. 
The gapful-gapless topological phase transition and associated evolution of Majorana states are demonstrated 
in a model for $D+p$-wave superconductivity. Experimental realization in recently fabricated cuprate heterostructures and heavy fermion thin films is discussed. 
\end{abstract}

\maketitle
\section{Introduction}
Topological superconductivity is one of the fascinating topics in modern condensed matter physics.\cite{Qi2011,Sato2016review,SatoAndo2016}
In particular, topologically nontrivial superconducting phases proposed in noncentrosymmetric systems attract much attention.\cite{Sato2009_STF,Sato2010_STF,Sau2010,Alicea2010,Lutchyn2010,Mourik2012,Nadj-Perge2014,Fu2008,Wang2012topo,Xu2014topo,Xu2015topo,Sun2016}
Inversion-symmetry breaking in crystal structures induces antisymmetric spin-orbit coupling (ASOC), and spin-momentum locking of electrons by ASOC causes unusual magnetic responses much different from centrosymmetric systems.
For example, $s$-wave Rashba superconductors (SCs) undergo topological transition at high magnetic fields, 
although centrosymmetric $s$-wave SCs are topologically trivial.\cite{Sato2009_STF,Sato2010_STF,Sau2010,Alicea2010,Lutchyn2010,Mourik2012,Nadj-Perge2014}
Further research of noncentrosymmetric SCs may identify a variety of topological phases which can not be achieved in centrosymmetric SCs.

Recently, a scenario for realizing gapful topological superconductors (TSCs) was proposed:\cite{Yoshida2016,Daido2016}
Two-dimensional (2D) {\it nodal} noncentrosymmetric SCs are gapped under the Zeeman field perpendicular to the momentum-dependent spin polarization axis by ASOC. They become 2D strong TSCs of symmetry class D\cite{Schnyder2008,Schnyder2009,Kitaev2009,Ryu2010,Morimoto2013,Samokhin2015}, when the spin-singlet-dominant Cooper pairing causes superconductivity. 
In sharp contrast to the topological $s$-wave superconductivity studied in the literature,\cite{Sato2009_STF,Sato2010_STF,Sau2010,Alicea2010,Lutchyn2010,Mourik2012,Nadj-Perge2014} any fine tuning of parameters is not required to realize the topological order. However, previous studies (\Refs{Daido2016}{Yoshida2016}) focused on the Zeeman field 
perpendicular to the ASOC, and the field-angle dependence has not been uncovered. 
We show that the Pauli-pair-breaking effect makes the excitation gapless under the tilted Zeeman field, and then 
the intriguing gapless topological superconducting phase is realized. 

In this paper, we study the topological phase transitions in the tilted Zeeman field 
and clarify the topological nature of each superconducting phase. 
First, we show the gapful-gapless transition in the bulk energy spectrum by tilting the Zeeman field. 
The stability of the gapful phase against a considerable tilting angle is revealed. 
Second, 
three kinds of topological edge states associated with topological phase transitions are elucidated: 
(1) Majorana flat bands in the absence of Zeeman field\cite{Yada2011,Sato2011,Schnyder2011}, 
(2) chiral Majorana edge states under the perpendicular Zeeman field\cite{Yoshida2016,Daido2016}, 
and (3) unidirectional Majorana edge states\cite{Wong2013,Baum2015} in the gapless phase under the tilted Zeeman field. 
Unidirectional Majorana edge states propagate in the same direction on both oppositely oriented edges, owing to the synergy of inversion-symmetry breaking and time-reversal-symmetry breaking.
Our arguments are based on the general Bogoliubov-de Gennes (BdG) Hamiltonian without assuming specific symmetry of the order parameter. 
Thus, the results obtained below are valid for most of the dominantly spin-singlet noncentrosymmetric 
superconducting states with nodal gap. The general results are demonstrated in a model of the $D+p$-wave Rashba SC
which may be realized in high-$T_{\rm c}$ cuprate heterostructures and heavy-fermion thin films.

The outline of this paper is as follows: In \Sec{sec:energyspectrum}, we calculate the energy spectrum of generic nodal noncentrosymmetric SCs under the low Zeeman field and illustrate the gapful-gapless transition in the tilted field.
In \Sec{sec:bulkexample}, we demonstrate the gapful-gapless transition in $D+p$-wave superconductivity.
{In \Sec{sec:Topology}, a unified view of topological phase transitions and associated topological edge states is given. In \Secs{subsec:MFB}{subsec:CME}, we review Majorana flat bands and chiral Majorana edge states, respectively. In \Sec{subsec:UME}, we show the emergence of unidirectional Majorana edge states from Majorana flat bands. These results are illustrated by the model for $D+p$-wave superconductivity. In \Sec{sec:ubiquitous}, the generality of the method for realizing unidirectional Majorana edge states is clarified. It is shown that we can choose a boundary direction hosting Majorana flat bands in noncentrosymmetric nodal SCs, regardless of the pairing symmetry of superconductivity. Then, the applied Zeeman field realizes unidirectional Majorana edge states. Finally, we give a brief summary of the paper in \Sec{sec:conclusion}. Heterostructures of high-temperature cuprate SCs and heavy fermion SCs are discussed as a platform of the topologically nontrivial gapful and gapless phases. }

\section{Energy spectrum of noncentrosymmetric nodal SC\MakeLowercase{s} under Zeeman field}
\label{sec:energyspectrum}
First, we introduce a general model for noncentrosymmetric SCs and discuss the excitation spectrum. 
We adopt a Bogoliubov-de Gennes (BdG) Hamiltonian
\begin{gather}
H_{\text{BdG}}(\bm{k})=\begin{pmatrix}
H_N(\bm{k})&\Delta(\bm{k})\\
\Delta(\bm{k})^\dagger&-H_N(-\bm{k})^T
\end{pmatrix},
\label{BdGHamiltonian}
\end{gather}
where $H_N(\bm{k})\equiv\xi(\bm{k})+\alpha\bm{g}(\bm{k})\cdot\bm{\sigma}-\muB\bm{H}\cdot\bm{\sigma}$ is the Hamiltonian in the normal state and $\Delta(\bm{k})\equiv(\psi(\bm{k})+\bm{d}(\bm{k})\cdot\bm{\sigma})i\sigma_y$ is the superconducting gap function. $\xi(\bm{k})$ is the kinetic energy of electrons measured from a chemical potential $\mu$, $\bm{g}(\bm{k})$ is the $g$ vector of ASOC, and $\bm{H}$ is the Zeeman field. The scalar order parameter of the spin-singlet component is denoted by $\psi(\bm{k})$, while $\bm{d}(\bm{k})$ is the vector order parameter of spin-triplet one. 
Both $\psi(\bm{k})$ and $\bm{d}(\bm{k})$ take finite value because of the parity-mixing due to inversion-symmetry breaking.\cite{Bauer2012}

The BdG Hamiltonian given by \Eq{BdGHamiltonian} assumes vanishing center-of-mass momentum of Cooper pairs. Strictly speaking, Cooper pairs may have finite center-of-mass momentum $\bm{q}_H$ in noncentrosymmetric SCs under Zeeman fields, referred to as helical superconductivity\cite{Barzykin2002,Dimitrova2003,Agterberg2007}, or Fulde-Ferrell-Larkin-Ovchinnikov superconductivity\cite{Fulde1964,Larkin1964}.
However, the induced momentum $\bm{q}_H$ is negligibly small under the low Zeeman field\cite{Yanase2007_Nonunitary} and, therefore, the following results obtained from \Eq{BdGHamiltonian} are not altered by introducing finite $\bm{q}_H$. 

\begin{figure}
  \centering
    \includegraphics[width=80mm]{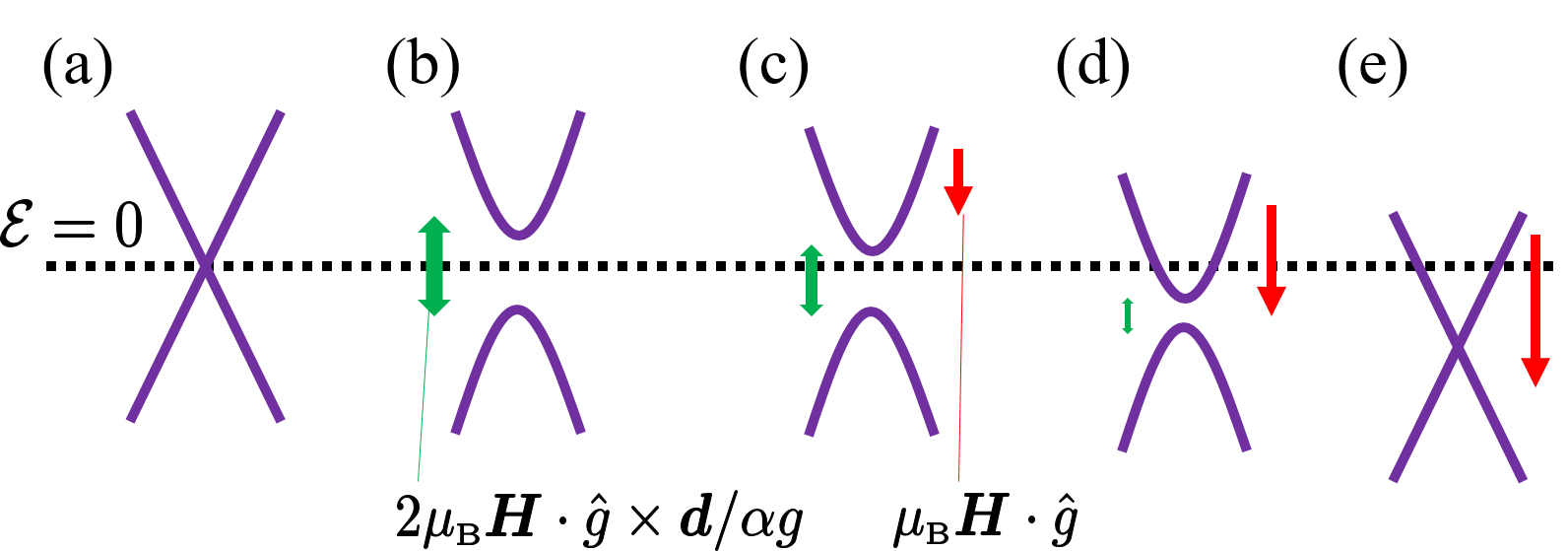}
    \caption{(Color Online) Illustration of gapful-gapless transition. We show energy spectrum around one of the nodal points on the $E_+$ Fermi surface, assuming $\muB\bm{H}\cdot\hat{g}\times\bm{d}/\alpha g\ge0$ and $\muB\bm{H}\cdot\hat{g}\ge0$ without loss of generality. (a) A point node in the absence of the Zeeman field, which is regarded as a massless Dirac cone. (b) The point node acquires a gap under the perpendicular field, leading to a massive Dirac cone. (c) Slightly tilted Zeeman field shifts the Dirac cone. The excitation remains gapful. (d) Gapless excitation spectrum under highly tilted Zeeman fields. (e) Spectrum under the parallel Zeeman field. The band gap closes.}
\label{fig0}
\end{figure}
\begin{figure}
  \centering
  \begin{tabular}{lll}
    (a)&(b)&(c)\\
    \includegraphics[width=27mm]{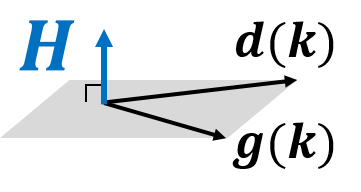}
&    \includegraphics[width=27mm]{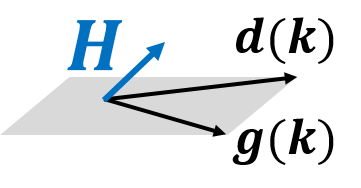}
&    \includegraphics[width=27mm]{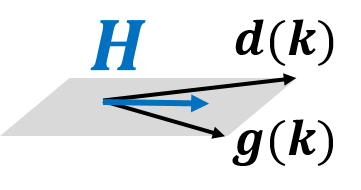}\\
  \end{tabular}
  \caption{Illustration of (a) perpendicular, (b) tilted, and (c) parallel Zeeman field. They are defined in terms of the plane spanned by $\bm{g}(\bm{k})$ and $\bm{d}(\bm{k})$, and may be different from the 2D plane of the system. In realistic SCs $\hat{g}(\bm{k})\times\bm{d}(\bm{k}) \ne 0$ is satisfied at general $\bm{k}$,\cite{Daido2016} and the above definition is well-defined.}
  \label{angle}
\end{figure}

In this paper, we consider weak-coupling SCs which preserve time-reversal symmetry unless the Zeeman field externally violates it. We ignore small modification of $\psi(\bm{k})$ and $\bm{d}(\bm{k})$ by low Zeeman fields, and then we can take $\psi(\bm{k})$ and $\bm{d}(\bm{k})$ to be real. The energy spectrum of the Bogoliubov quasiparticle under low Zeeman fields is given by \Ref{Daido2016}:
\begin{align}
\mathcal{E}_+&=-\muB\bm{H}\cdot\hat{g}\notag\\
&\pm\sqrt{E_+^2+\bigl|\psi+\bm{d}\cdot\hat{g}\bigr|^2+\bigl|\muB\bm{H}\cdot\hat{g}\times\bm{d}/\alpha g\bigr|^2},\label{lambda+}\\
\mathcal{E}_-&=+\muB\bm{H}\cdot\hat{g}\notag\\
&\pm\sqrt{E_-^2+\bigl|\psi-\bm{d}\cdot\hat{g}\bigr|^2+\bigl|\muB\bm{H}\cdot\hat{g}\times\bm{d}/\alpha g\bigr|^2},\label{lambda-}
\end{align}
where $E_\pm(\bm{k})\equiv\xi(\bm{k})\pm\alpha g(\bm{k})=0$ gives the two Fermi surfaces (FSs) split by ASOC. 
$g(\bm{k})\equiv|\bm{g}(\bm{k})|$ and $\hat{g}(\bm{k})\equiv\bm{g}(\bm{k})/g(\bm{k})$ are amplitude and direction of 
the $g$ vector, respectively. {Here the argument of $\mathcal{E}_\pm(\bm{k})$, $E_\pm(\bm{k})$, $\psi(\bm{k})$, $\bm{d}(\bm{k})$, $g(\bm{k})$, and $\hat{g}(\bm{k})$ is omitted, for simplicity.}
Equations \eqref{lambda+} and \eqref{lambda-} are valid when the conditions
\begin{gather}
\muB H\ll\alpha g(\bm{k}),\label{lowmagcond1}\\
|\psi(\bm{k})|\ll\alpha g(\bm{k}),\label{lowmagcond2}\\
d(\bm{k})\ll\alpha g(\bm{k}),\label{lowmagcond3}
\end{gather}
are satisfied, where we denoted $H\equiv|\bm{H}|$ and $d(\bm{k})\equiv |\bm{d}(\bm{k})|$.
Hereafter, we focus on the low Zeeman field satisfying \Eq{lowmagcond1} around originally nodal points.

It is indicated from Eqs.~\eqref{lambda+} and \eqref{lambda-} that the cooperation of inversion-symmetry breaking and time-reversal-symmetry breaking effectively realizes chiral superconductivity. 
The imaginary part of the gap function $\muB\bm{H}\cdot\hat{g}\times\bm{d}/\alpha g$ is given by the scalar product of the Zeeman field, $g$ vector of ASOC, and vector order parameter. 
In order to see the effects of ASOC and the Zeeman field, let us focus on the low energy spectrum. The nodal point $\bm{k}_0$ at $H=0$ is given by $E_\pm(\bm{k}_0)=\psi(\bm{k}_0)\pm\bm{d}(\bm{k}_0)\cdot\hat{g}(\bm{k}_0)=0$, leading to the massless Dirac cone illustrated in \Fig{fig0}(a). 
When the Zeeman field is applied perpendicular to the plane spanned by $\bm{d}(\bm{k}_0)$ and $\bm{g}(\bm{k}_0)$ 
[see \Fig{angle}(a)], the energy spectrum is gapped out as illustrated in \Fig{fig0}(b). 
The induced energy gap is given by the mass term $\mu_B\bm{H}\cdot\hat{g}\times\bm{d}/\alpha g$. 
When we slightly tilt the field as in \Fig{angle}(b), the spectrum shifts owing to the paramagnetic term 
$\pm \muB\bm{H}\cdot\hat{g}$. However, the spectrum remains gapful as long as 
\begin{gather}
|\muB\bm{H}\cdot\hat{g}|<|\muB\bm{H}\cdot\hat{g}\times\bm{d}/\alpha g|,
\label{gapfulcond}
\end{gather}
is satisfied around $\bm{k}_0$ [\Fig{fig0}(c)].
Further tilting violates the condition (\ref{gapfulcond}) and makes the excitation gapless, as shown 
in \Fig{fig0}(d). Finally, the parallel Zeeman field in the plane of $\bm{d}(\bm{k}_0)$ and $\bm{g}(\bm{k}_0)$ 
[\Fig{angle}(c)] closes the band gap at $\bm{k}=\bm{k}_0$ [\Fig{fig0}(e)].
\begin{figure}
\centering
\includegraphics[width=60mm]{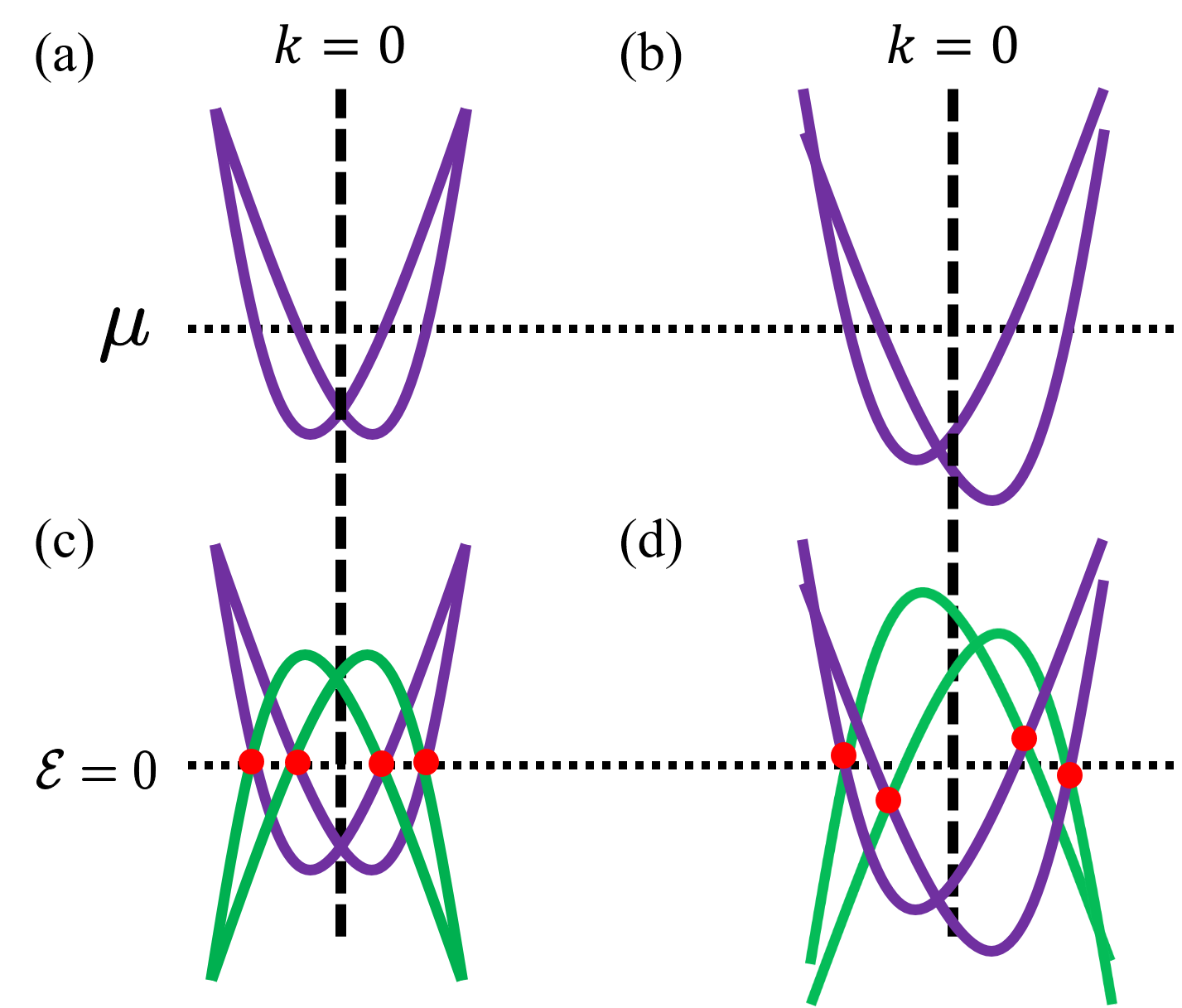}
\caption{(Color Online) Schematic figures of the paramagnetic shift of Dirac points. (a) Energy bands of electrons in the absence of Zeeman field. (b) Energy bands under the parallel Zeeman field. We choose the horizontal $k$-axis so that $\bm{H}\parallel\hat{g}(k)$. Bogoliubov quasiparticle spectrum (c) in the absence of Zeeman field and (d) under the parallel Zeeman field. 
We show the Dirac points by red circles. 
}
\label{fig01}
\end{figure}




\begin{figure*}[htbp]
  \centering
  \begin{tabular}{llll}
    (a) $\muB H=0$&(b) $\theta=0$ &(c) $\theta=\pi/4$&(d)\\
    \includegraphics[width=40mm]{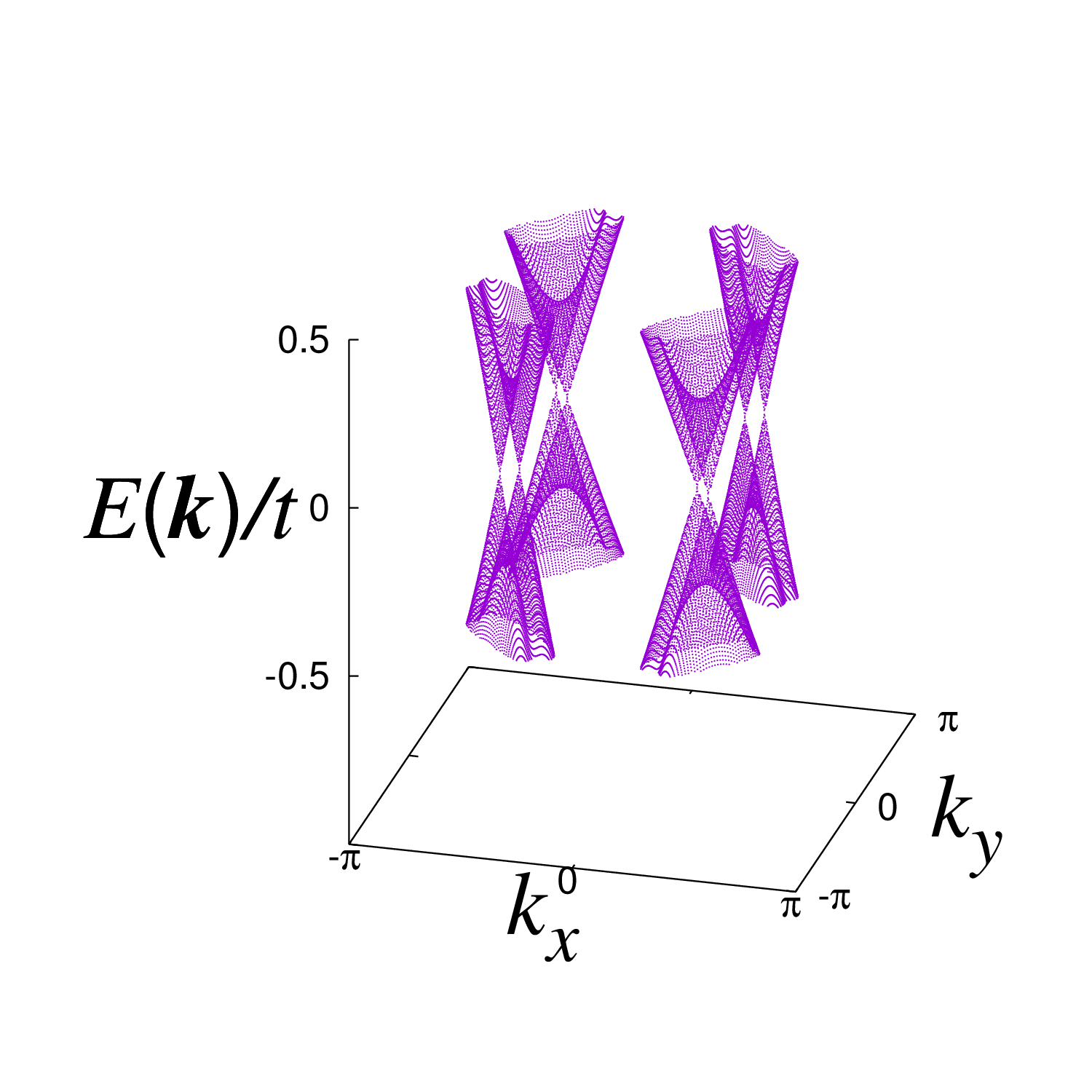}&\includegraphics[width=40mm]{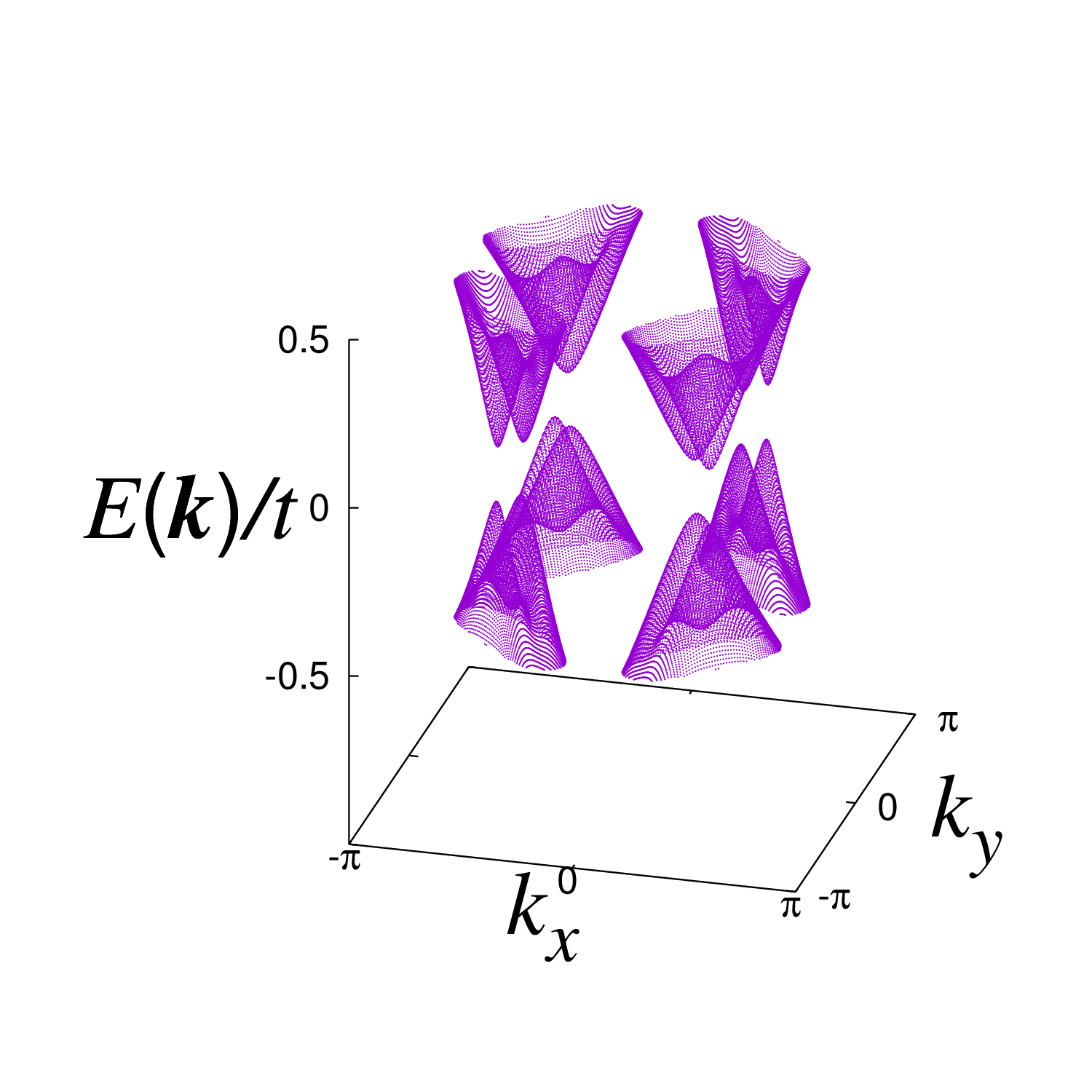}&\includegraphics[width=40mm]{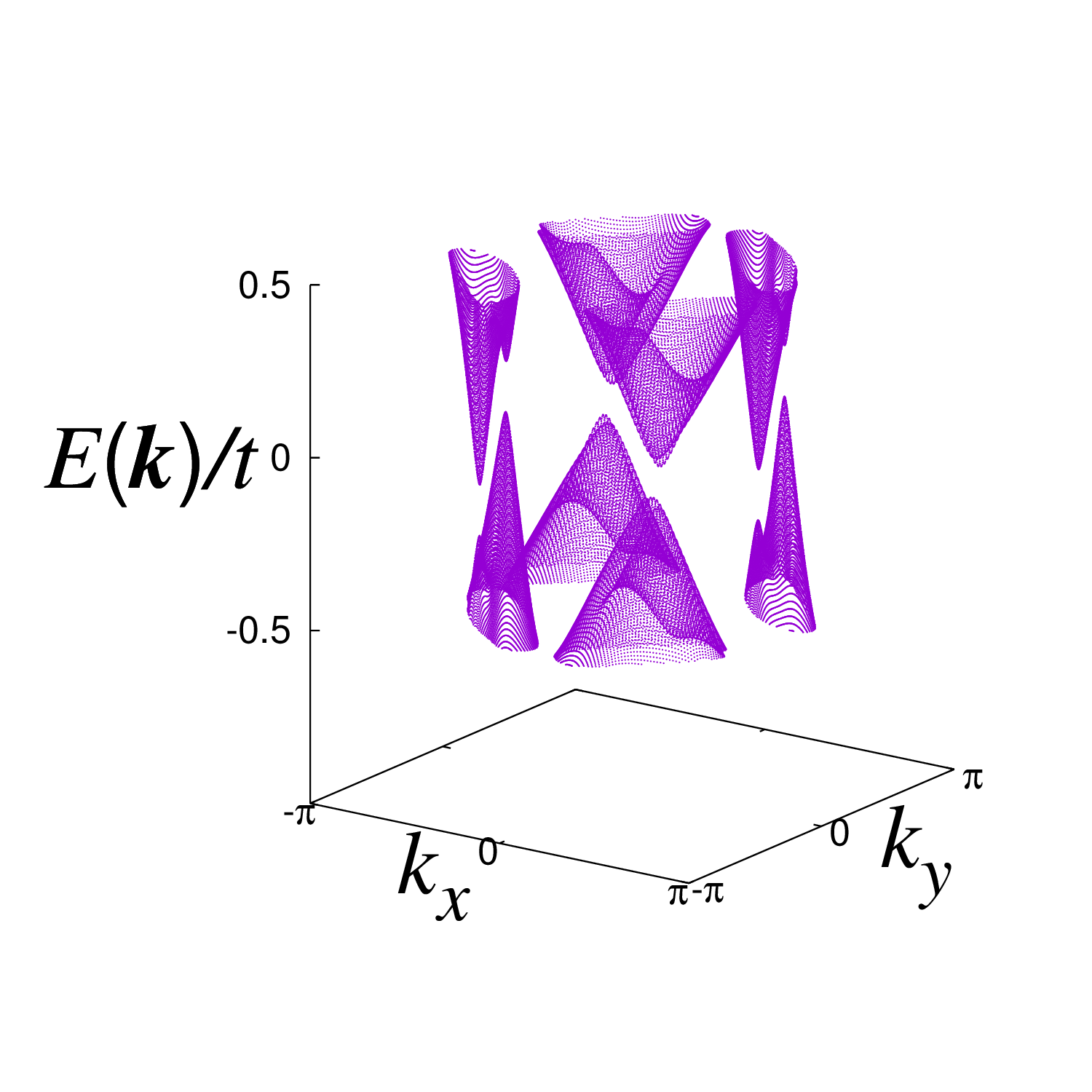}&\includegraphics[height=40mm]{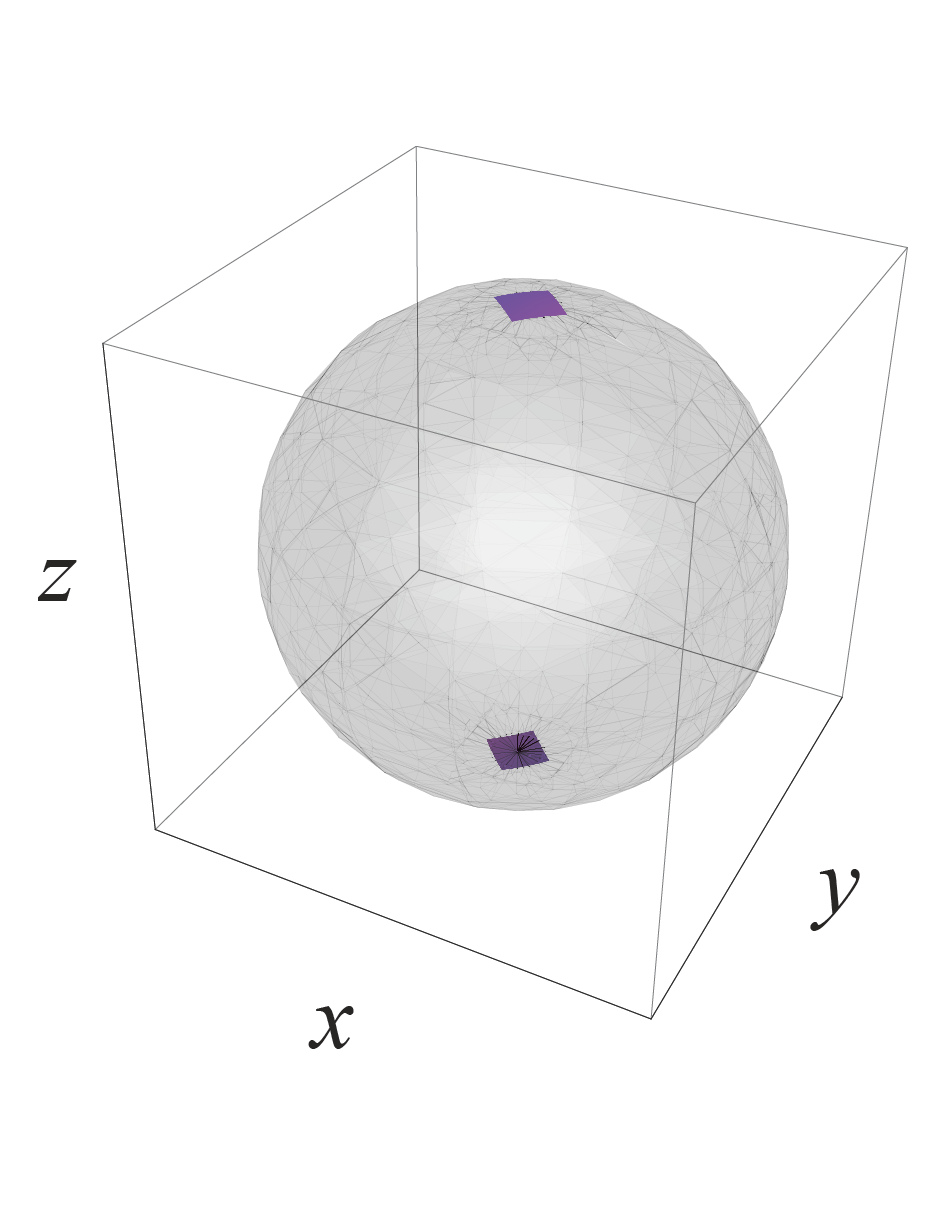}
  \end{tabular}
  \caption{(Color online) Bulk energy spectrum of a $D+p$-wave SC (a) in the absence of Zeeman field, and (b,c) under Zeeman field $\muB H=0.3$. (b) The Zeeman field is perpendicular to $\bm{g}(\bm{k})$ and $\bm{d}(\bm{k})$ ($\theta=0$). (c) Zeeman field is tilted and we take $\theta=\pi/4$ and $\phi=3\pi/4$. 
We choose $t=1$, $t'=0.2$, $\mu=-0.79$, $\alpha=0.3$, $\psi_0=0.5$, and $d_0=0.1$. (Comparably large order parameters and Zeeman field are adopted for visibility.)
    (d) The angle of Zeeman field is represented by a sphere. The gapful (gapless) topological superconducting phase is 
shown by purple (gray). We use \Eq{gapful_region} for $d_0/\alpha=0.1$.
  }
\label{bulkband}
\end{figure*}

The response to the Zeeman field may differ from nodes to nodes. Therefore, global excitation gap $\Delta E$ 
is given by the minimum of the local energy gap among massive Dirac cones at originally nodal points. 
Gapful-gapless transition occurs when one of the massive Dirac cones crosses the Fermi level and $\Delta E$ vanishes.

Note that both inversion-symmetry breaking and time-reversal-symmetry breaking are essential for the shift 
of the Dirac spectrum from $\mathcal{E}=0$. Otherwise, the BdG Hamiltonian has a symmetric spectrum with respect to 
$\mathcal{E}=0$, and then the center of the Dirac cones is pinned to $\mathcal{E}=0$.
For an intuitive understanding of this point, we show schematic figures in \Fig{fig01}. 
Figure~\ref{fig01}(a) illustrates the dispersion relation $E_\pm$ on a certain line 
in the Brillouin zone (BZ) respecting the particle-hole symmetry. 
The spin degeneracy is lifted except for at the time-reversal-invariant momentum, and spins are aligned in each band 
along the direction $\pm\hat{g}$, as a result of inversion-symmetry breaking. 
When the Zeeman field is applied parallel to the $g$ vector, the $E_\pm$-band is deformed asymmetrically in terms of $k=0$ 
(time-reversal-invariant momenta) [\Fig{fig01}(b)]. 
The Bogoliubov quasiparticle spectrum is shown in \Figss{fig01}(c) and \ref{fig01}(d). 
Around the nodal points where $\Delta(\bm{k}) = 0$, the quasiparticle spectrum is described by the set of 
$E_\pm(\bm{k})$ and $-E_\pm(-\bm{k})$ (which may be modified by the Zeeman field) owing to the particle-hole symmetry. 
The Dirac points are pinned to the zero energy $\mathcal{E}=0$ at $H=0$ [\Fig{fig01}(c)], while 
Dirac cones are shifted by the Zeeman field [\Fig{fig01}(d)]. 
When $\Delta(\bm{k})\neq0$, the Dirac spectrum may be gapped. 

\section{Illustration of gapful-gapless transition in $D+p$-wave SC}
\label{sec:bulkexample}

So far we have not assumed any specific symmetry of superconductivity. 
In this section, we demonstrate the gapful-gapless transition illustrated in Sec.~II 
through the analysis of a model for $D+p$-wave superconductivity. 
The critical angle of the Zeeman field for the gapful-gapless transition is estimated, and the stability 
of the gapful TSCs against experimentally unavoidable field-angle fluctuations is shown. 

We consider a 2D $D+p$-wave SC, taking
\begin{gather}
\xi(\bm{k})=-2t(\cos k_x+\cos k_y)+4t'\cos k_x\cos k_y-\mu,\label{D+pmodel1}\\
\bm{H}=H(\sin\theta\cos\phi,\,\sin\theta\sin\phi,\,\cos\theta)^T,\\
\bm{g}(\bm{k})=(-\sin k_y,\,\sin k_x,\,0)^T,\label{Rashba}\\
\psi(\bm{k})=\psi_0(\cos k_x -\cos k_y),\\
\bm{d}(\bm{k})=d_0(\sin k_y,\, \sin k_x,\,0)^T.\label{D+pmodel2}
\end{gather} 
In the absence of the Zeeman field, the bulk spectrum has eight excitation nodes at intersections of the diagonal directions $k_x=\pm k_y$ and the two FSs split by the ASOC [\Fig{bulkband}(a)].

Since the model includes the Rashba-type ASOC, the perpendicular direction defined by Fig.~\ref{angle}(a) 
is along the $c$-axis ($\theta=0$). Under the Zeeman field in the perpendicular direction, 
we have $\bm{H}\cdot\hat{g}=0$ and obtain the gapful bulk spectrum in \Fig{bulkband}(b). 
The excitation gap is estimated from \Eqs{lambda+}{lambda-}, 
\begin{gather}
\bigl|\muB\bm{H}\cdot\hat{g}\times\bm{d}/\alpha g\,\bigr|_{\text{nodes}}=d_0\frac{\muB H}{\alpha}\quad(\theta=0).
\end{gather}

For the Zeeman field tilted from the perpendicular direction, the paramagnetic term
\begin{gather}
\muB\bm{H}\cdot\hat{g}\bigr|_{\text{nodes}}=\pm\muB H\sin\theta\sin(\phi\pm\pi/4)
\end{gather}
in \Eqs{lambda+}{lambda-} shifts the Dirac cones, and the mass term reduces to
\begin{equation*}
\bigl|\muB\bm{H}\cdot\hat{g}\times\bm{d}/\alpha g\,\bigr|_{\text{nodes}}=d_0\frac{\muB H|\cos{\theta|}}{\alpha}. 
\end{equation*}
The competition between those two terms determines the gapful-gapless transition. 
The global excitation gap $\Delta E$ is obtained as 
\begin{align}
  \Delta E&=\max\Bigl[0,\ \min_{\text{nodes}}\bigl\{-\left|\mu_B\bm{H}\cdot\hat{g}\right|+\left|\mu_B\bm{H}\cdot\hat{g}\times\bm{d}/\alpha g\right|\}\Bigr],\\
  &=\max\Bigl[0,\ \min_{\pm}\bigl\{-\muB H\left|\sin\theta\sin(\phi\mp\pi/4)\right|\notag\\
    &\quad\quad\quad\quad\qquad\qquad+d_0\muB H|\cos\theta|/\alpha\bigr\}\Bigr].
\label{energy_gap}
\end{align}
Equation \eqref{energy_gap} ensures that $\Delta E$ is finite in a certain angle region around $\theta=0$ and $\theta=\pi$ [see \Fig{bulkband}(d)], indicating the gapful superconducting state. The gapful state changes to be gapless by tilting the Zeeman field from $\theta=0$ toward $\theta=\pi/2$.
Figure~\ref{bulkband}(c) shows typical bulk bands in the gapless phase. 
Note that we choose $\phi=3\pi/4$ and hence, the paramagnetic term vanishes at nodal points on the $k_x=-k_y$ line. 

The critical angle of the gapful-gapless transition $\theta_c(\phi)$ is given by
\begin{equation}
\frac{d_0}{\alpha}=|\tan\theta_c(\phi)|\max_\pm|\sin(\phi\pm\pi/4)|.
\label{gapful_region}
\end{equation}
Equation \eqref{gapful_region} approximately consists of four polar equations whose solution is a straight line, 
and therefore, the gapful region is two nearly square sheets around $\theta=0$ and $\theta=\pi$ [Fig. \ref{bulkband}(d)].
The critical angle $\theta_c$ is roughly estimated to be as large as \SI{6}{\degree} in cuprate heterostructures, 
by adopting $d_0/\alpha\sim\psi_0/E_{\rm F}\sim 1/10$.\cite{Fujimoto2007review,Yanase2003} 
Thus, a large critical angle is obtained in high-temperature superconductors with large $T_{\rm c}/E_{\rm F}$ 
even when the spin-orbit coupling is small. 
The obtained value \SI{6}{\degree} is much larger than the experimental uncertainty of the field angle. 
Thus, we conclude that the paramagnetically induced excitation gap is robust against angle fluctuations. 
Since this gapful superconducting state is known to be topologically nontrivial and specified 
by the Chern number $-4$,\cite{Yoshida2016,Daido2016} it is also concluded that the gapful TSCs are robust against field-angle fluctuations. 


\section{Unified view of topological edge states in noncentrosymmetric SC\MakeLowercase{s}}
\label{sec:Topology}
In the following part of this paper, we investigate the edge states in noncentrosymmetric SCs described 
by the Hamiltonian given by \Eq{BdGHamiltonian}.
In particular, we clarify the relationship between a variety of topological edge states 
under zero and low Zeeman fields. A unified view on the Majorana flat band, chiral Majorana edge state, 
and unidirectional Majorana edge state is discussed. 

\subsection{Majorana flat bands in time-reversal-symmetric phase}
\label{subsec:MFB}
First, we review the well-known Majorana flat bands in the time-reversal symmetric nodal SCs\cite{Yada2011,Sato2011,Schnyder2011} from the perspective of the formulas obtained in \Ref{Daido2016}.
Topological invariants of the nodal (weak) topological SCs may be given by the one-dimensional (1D) winding number of class AI\hspace{-.1em}I\hspace{-.1em}I:\cite{Yada2011,Sato2011,Schnyder2011}
\begin{gather}
  W(k_x)\equiv-\int^\pi_{-\pi}\frac{dk_y}{{4\pi i}}\,\tr\left[\Gamma H_{\text{BdG}}(\bm{k})^{-1}\frac{\partial}{\partial {k_y}}H_{\text{BdG}}(\bm{k})\right],
\label{1DwindGamma}
\end{gather}
where $\Gamma$ is a chiral operator and the chiral symmetry of $H_{\text{BdG}}$ is expressed as:
\begin{gather}
\Gamma\equiv\begin{pmatrix}0&\sigma_y\\\sigma_y&0\end{pmatrix},
\quad\{\Gamma,H_{\text{BdG}}(\bm{k})\}=0.
\end{gather}
The chiral symmetry is obtained by combining particle-hole symmetry with time-reversal symmetry.
Below, we show a compact formula for the winding number $W(k_x)$. 

{Note that the winding number is antisymmetric with respect to $k_x$, because of the time-reversal symmetry\cite{Yuansen2017}:
\begin{gather}W(k_x)=-W(-k_x).\end{gather}
In particular, the winding number vanishes at time-reversal-invariant momenta.
Thus, we obtain $W(-\pi)=0$, when the bulk spectrum at $k_x=-\pi$ is gapful.}


In previous studies, \Refs{Yada2011}{Sato2011}, the winding number has been estimated by numerical calculations 
or by using a formula requiring us to trace sign changes of the energy and order parameter along the $k_y$ line. 
However, the winding number can be obtained by a more compact formula: all we have to do is to calculate the
``winding number of nodes.''


In the absence of the Zeeman field, nodal points are indeed protected by the winding number defined in a similar form,\cite{Schnyder2011}
\begin{gather}
    W_\pm(\bm{k}_0)\equiv-\oint_{C_\pm(\bm{k}_0)}\frac{d\bm{k}}{4\pi i}\,\cdot\tr\left[\Gamma H_{\text{BdG}}(\bm{k})^{-1}\nabla_{\bm{k}}H_{\text{BdG}}(\bm{k})\right],
\label{winddef}
\end{gather}
where $C_\pm(\bm{k}_0)$ is a sufficiently small loop running anticlockwise around the nodal point $\bm{k}_0$ on the FS of the $E_\pm$ band.
The winding number of nodes given by \Eq{winddef}, is expressed by a simple formula\cite{Daido2016},
\begin{gather}
W_\pm(\bm{k}_0)=-\sgn\left[\partial(\psi\pm\bm{d}\cdot\hat{g})/\partial k_\parallel\right]_{\bm{k}_0},\label{formulawind2}
\end{gather}
where $k_\parallel$ is a coordinate taken parallel to the $E_\pm$ FS viewing the $E_\pm>0$ region in the right-hand side,
\begin{gather}
\partial/\partial k_\parallel\equiv\hat{k}_\parallel\cdot\nabla_{\bm{k}},\\
\hat{k}_\parallel\equiv\frac{\hat{z}\times\nabla_{\bm{k}} E_\pm(\bm{k})}{|\hat{z}\times\nabla_{\bm{k}} E_\pm(\bm{k})|}.
\end{gather}
Now the winding number $W(k_x)$ can be evaluated by counting the winding number of nodes. 
The periodicity of the BZ allows us to form a loop by combining the integration line $k_x=\tilde{k}_x$ with the line $k_x=-\pi$. Thus, we obtain 
\begin{align}
  W(\tilde{k}_x)&=W(-\pi)+\sum_{\substack{(\pm,\bm{k}_0);\\\text{encircled}}}W_\pm(\bm{k}_0)\\
  &=\sum_{\substack{(\pm,\bm{k}_0);\\\text{encircled}}}W_\pm(\bm{k}_0)\label{formulawind}, 
\end{align}
{where the summation is taken over the nodes encircled by the loop, that is, nodes in the domain $-\pi<k_x<\tilde{k}_x$.\cite{Samokhin2016}}
The compact formula given by \Eq{formulawind} is valid when the bulk excitation is gapped at $k_x=-\pi$.
{It is straightforward to derive similar formulas taking other time-reversal invariant momenta (TRIM) instead of $k_x=-\pi$ in such a situation.}

The formula \eqref{formulawind2} is furthermore simplified in the usual cases. To be specific, we consider spin-singlet superconductivity slightly admixed with spin-triplet superconductivity. (Results for spin-triplet-dominant SCs are easily reproduced by replacing $\psi$ by $\pm\bm{d}\cdot\hat{g}$ in the following results.)
Then, we obtain
\begin{gather}
  W_\pm(\bm{k}_0)=-\sgn[\partial \psi/\partial k_\parallel]_{\bm{k}_0}. 
\label{singletwind}
\end{gather}
Equation~\eqref{singletwind} reveals that the winding number of nodes is determined only by the gradient of 
$\psi$ at the node. 

\begin{figure}
  \includegraphics[width=80mm]{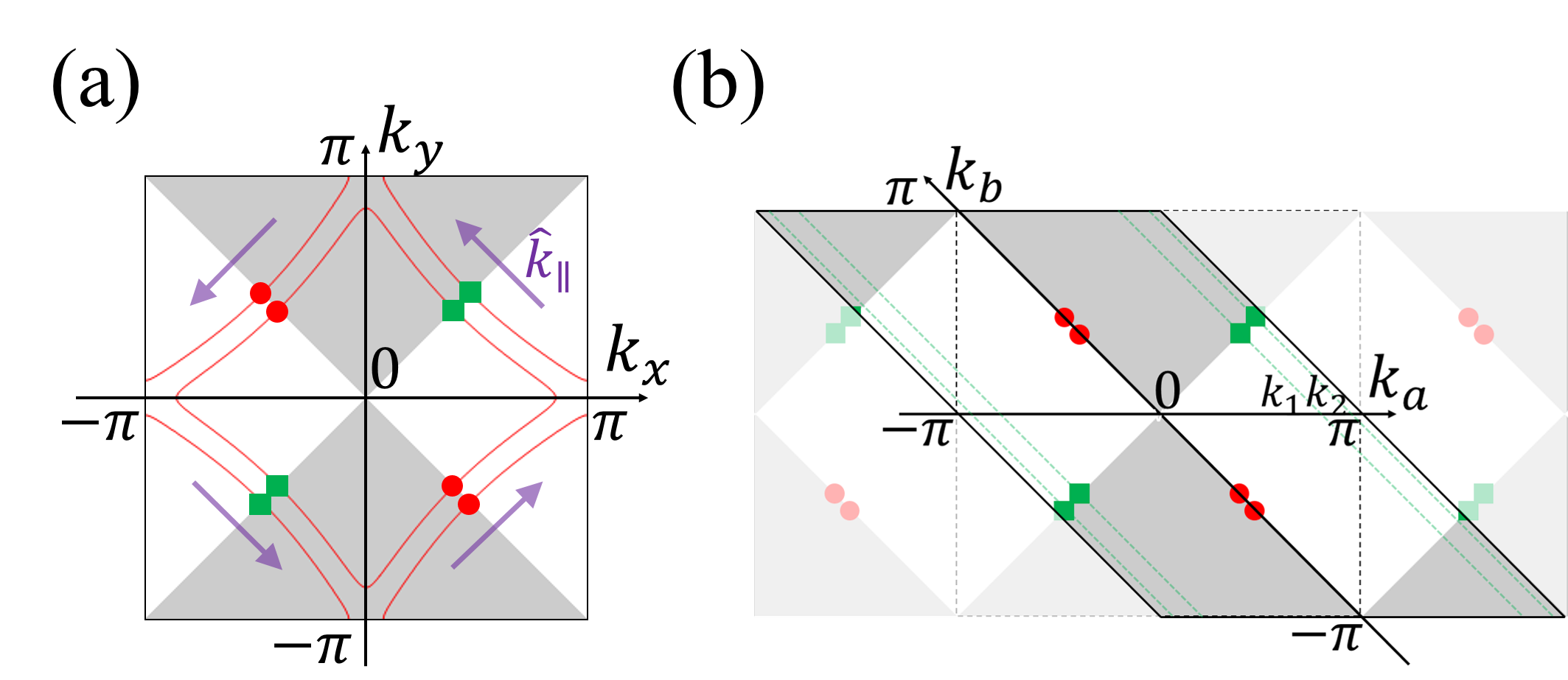}
  \caption{(Color Online) FSs and nodal points of a $d_{x^2-y^2}$-wave SC in the BZ for crystal translation vectors (a) $\Set{\hat{x},\hat{y}}$, and (b) $\Set{\bm{a},\bm{b}}$. Red lines in (a) show FSs for $\mu=-0.79$ in the model for $D+p$-wave SC (\Sec{sec:bulkexample}). Red circles and green squares show the nodes with winding number $+1$ and $-1$, respectively. In the shaded region $\psi(\bm{k})>0$. Purple arrows indicate the direction of $\hat{k}_\parallel$.}
  \label{exwind}
\end{figure}

To show the usefulness of the formula, we consider the $d_{x^2-y^2}$ superconductivity with the gap function 
$\psi(\bm{k})=|\psi_0|(\cos k_x-\cos k_y)$ and the Rashba-type ASOC in \Eq{Rashba}. 
Nodes on diagonal axes are characterized by the winding number $+1$ or $-1$, as shown in \Fig{exwind}(a) 
by red circles and green squares, respectively.
The winding number of nodes is $-1$ (green squares) in the first quadrant, since the sign of $\psi$ changes 
from negative (white region) to positive (shaded region) along the $\hat{k}_\parallel$ direction. 
Since the order parameter $\psi$ is $C_4$-odd, the winding number of nodes is $+1$, $-1$, and $+1$ 
in the second, third, and fourth quadrant, respectively. 
In this way, we can easily attribute winding number $\pm1$ to nodes in general.

\begin{figure}
  \begin{tabular}{ll}
    (a) $\muB H=0$ &(b) $\muB H=0$ \\
    \includegraphics[width=40mm]{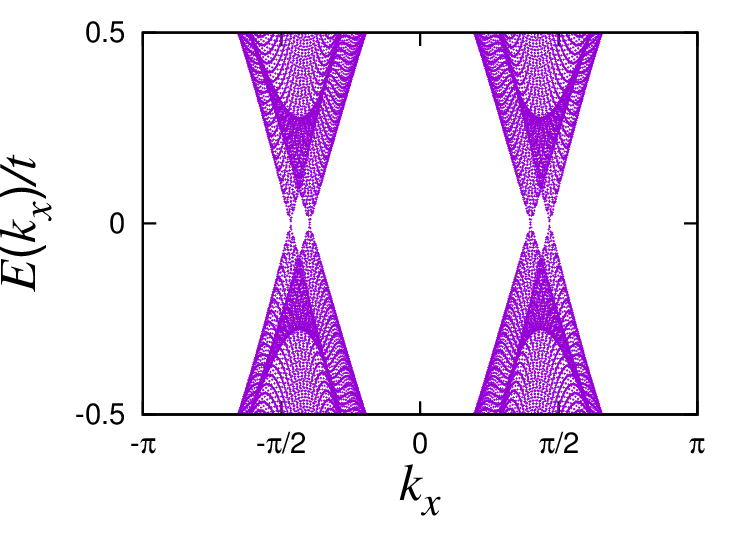}&\includegraphics[width=40mm]{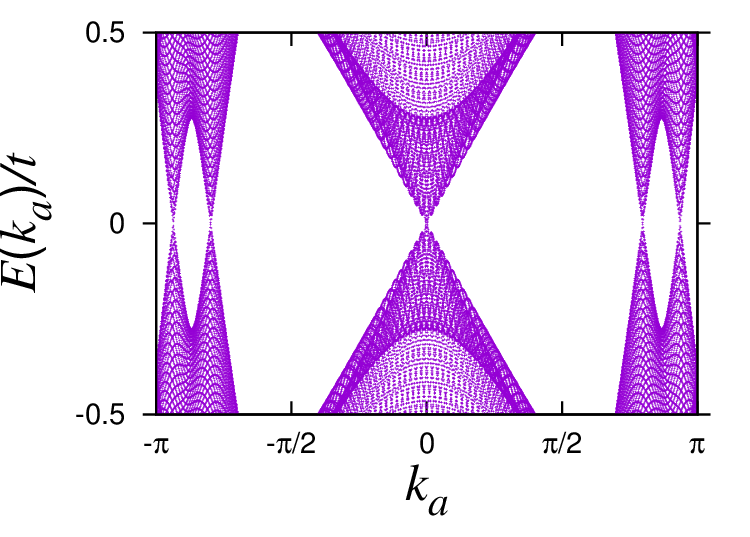}\\
    (c) $\muB H=0$ &(d) $\muB H=0$ \\
    \includegraphics[width=40mm]{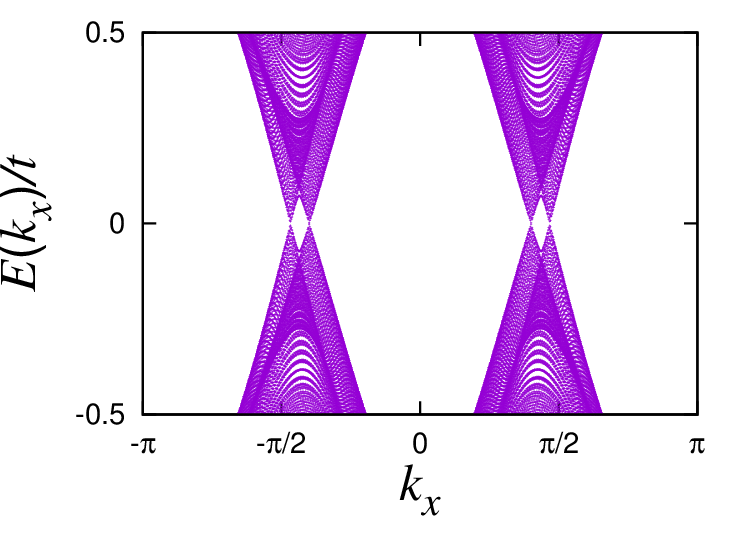}&\includegraphics[width=40mm]{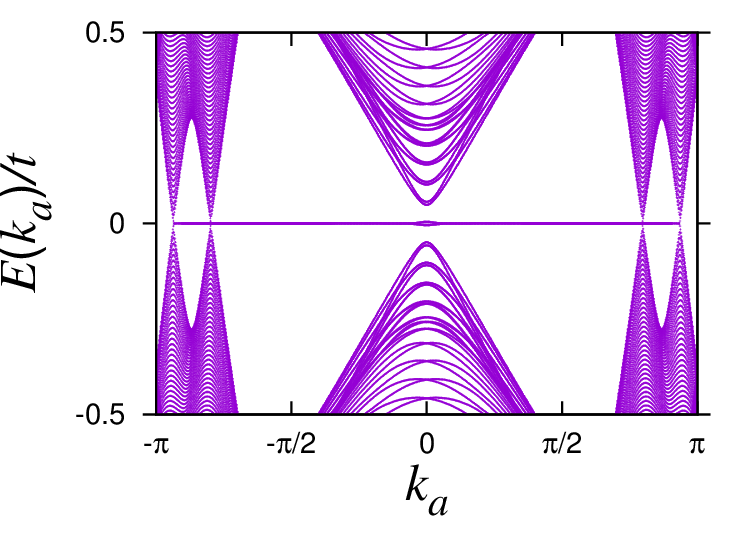}
  \end{tabular}
  \caption{Energy spectrum of a $D+p$-wave SC in the absence of Zeeman field.
Parameters used are the same as \Fig{bulkband}(a).
    (a,b) Bulk energy spectrum projected onto the edge BZ for (a) the $(010)$ edge and (b) the $(1\bar{1}0)$ edge. 
(c,d) Spectrum of ribbon-shaped systems for open boundary condition along (c) the $(010)$ edge and (d) the $(1\bar{1}0)$ edge. No edge states appear on the $(010)$ edge, while Majorana flat bands appear on the $(1\bar{1}0)$ edge at momenta $|k_a| \lesssim \pi$. Note that the bulk spectrum around $k_a\simeq0$ is gapless although a small gap is induced by the finite-size effect. We take $200$ lattice sites in the $\bm{b}$ direction.}
    \label{flatbands}
\end{figure}

Now the winding number $W(k_x)$ is evaluated from the formula \eqref{formulawind}. 
In the $d_{x^2-y^2}$-wave SC, nodes with winding number $\pm 1$ form a pair on a line along the $k_y$ axis. 
Therefore, contributions to $W(k_x)$ are completely canceled out. Thus, we obtain $W(k_x)=0$. 
In accordance with the bulk-edge correspondence, the edge state does not appear on the $(010)$ edge as shown in 
\Fig{flatbands}(c). 

\begin{figure}
  \centering
  \includegraphics[height=33mm]{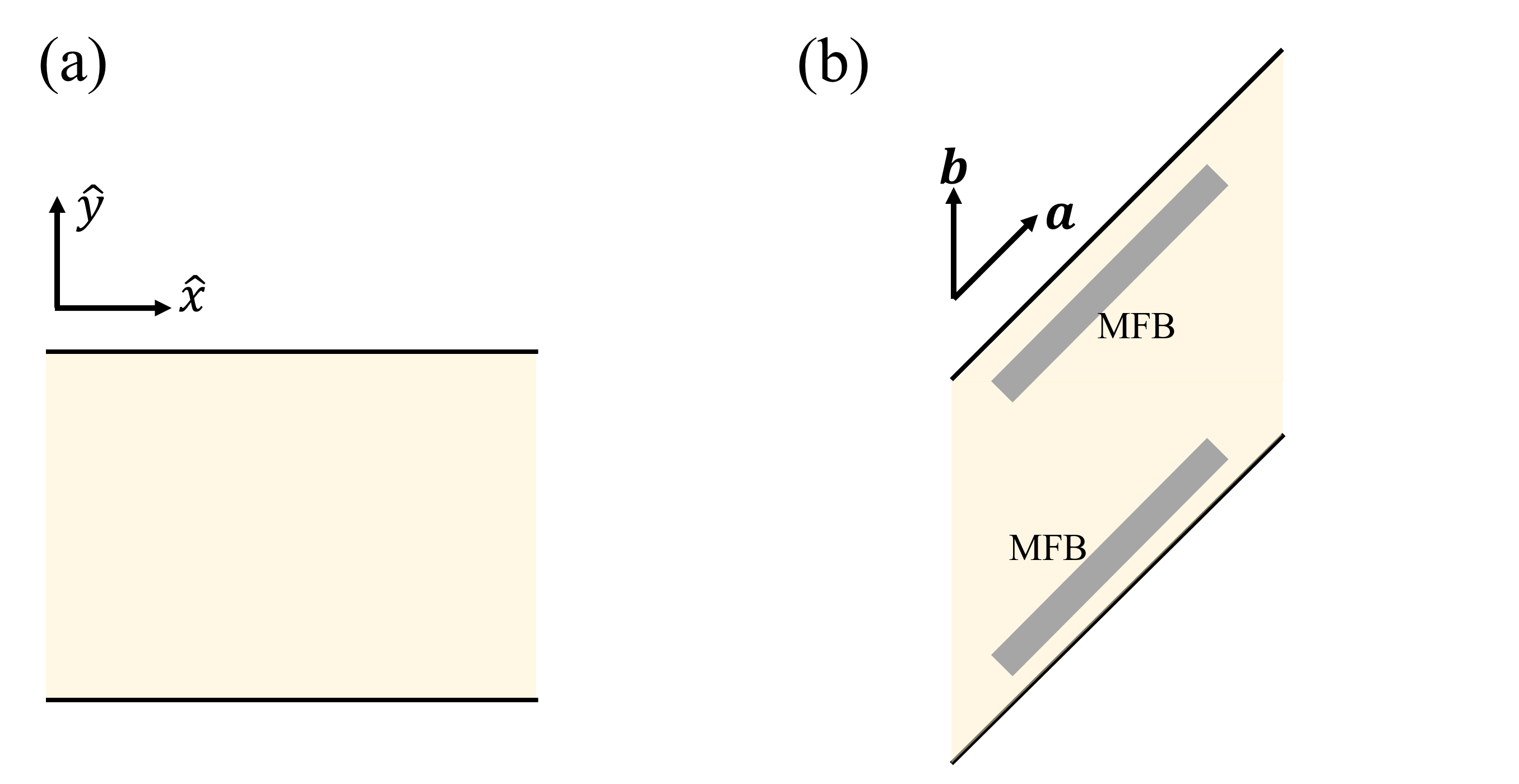}
  \caption{(Color Online) Illustration of boundary conditions and crystal translation vectors for (a) $\Set{\hat{x},\hat{y}}$ and (b) $\Set{\bm{a},\bm{b}}$. Black lines represent the open boundary. Corresponding to numerical results in \Fig{flatbands}, we show Majorana flat bands by gray lines on the $(1\bar{1}0)$ and $(\bar{1}10)$ edges in (b). 
}
  \label{crystalvectors}
\end{figure}

The situation is totally different on the $(1\bar{1}0)$ edge. 
To analyze the topological edge state, 
let us change the crystal translation vectors as
\begin{gather}
\Set{\hat{x},\,\hat{y}}\to\Set{\bm{a}\equiv\hat{x}+\hat{y},\,\bm{b}\equiv\hat{y}},\label{tiltedset}
\end{gather}
and consider the open (periodic) boundary condition in the $\bm{b}$ ($\bm{a}$) direction [\Fig{crystalvectors}(b)]. 
Reciprocal lattice vectors corresponding to $\Set{\bm{a},\,\bm{b}}$ are given by
\begin{gather}
  \Set{\bm{\alpha}\equiv\hat{x},\,\bm{\beta}\equiv-\hat{x}+\hat{y}},
\end{gather} and wave numbers in the bulk are represented as 
\begin{gather}
  \bm{k}=k_a\bm{\alpha}+k_b\bm{\beta},\\
  k_a\equiv k_x+k_y,\quad k_b\equiv k_y. 
\end{gather}
Only $k_a$ is a good quantum number for the ribbon-shaped boundary condition. 

The winding number $W(k_a)$ is defined in the same way as \Eq{1DwindGamma}, and 
the integration path should be taken parallel to the $k_b$ axis. 
Figure~\ref{exwind}(b) shows that the nodal points projected to the $k_a$ axis are 
placed at $k_a=0$, $\pm k_1,$ and $\pm k_2$.
Then, the formula (\ref{formulawind}) immediately reads
\begin{gather}
  W(k_a)=\begin{cases}
  0\quad(-\pi<k_a<-k_2)\\
  -1\quad(-k_2<k_a<-k_1)\\
  -2\quad(-k_1<k_a<0)\\
  2\quad(0<k_a<k_1)\\
  1\quad(k_1<k_a<k_2)\\
  0\quad(k_2<k_a\le\pi)
  \end{cases},
\end{gather}
which is consistent with the result for $d_{xy}+p$-wave superconductivity,\cite{Sato2011} as expected. 
According to the bulk-edge correspondence, the finite winding number ensures the presence of 
Majorana flat bands\cite{Sato2011}. 
Indeed, \Fig{flatbands}(d) shows the Majorana flat bands on the $(1\bar{1}0)$ edge.


Summarizing, the time-reversal-invariant nodal noncentrosymmetric SCs are weak TSCs characterized by 
the 1D winding number of class AIII. 
The Majorana flat band may appear in the edge state, depending on the boundary direction. 
The condition for the Majorana flat band is immediately understood by the compact formula for 
the winding number given by \Eq{formulawind}. 

\subsection{Chiral Majorana edge states in gapful topological superconducting phase}
\label{subsec:CME}
Now we turn to the time-reversal-symmetry-broken state by the Zeeman field. 
As we have already mentioned in Sec.~II, the energy spectrum is gapful under the condition given by \Eq{gapfulcond}. 
Gapful SCs without time-reversal symmetry belong to the symmetry class D, and thus 
the topological invariant is the Chern number defined by\cite{Thouless1982}
\begin{gather}
\nu\equiv\int\frac{d^2k}{2\pi i}\sum_{i,j;\ n\in P}{\epsilon_{ij}}\frac{\partial}{\partial k_i}\braket{u_n(\bm{k})|\frac{\partial}{\partial k_j}|u_n(\bm{k})},\label{chern}
\end{gather}
where {$\epsilon_{ij}$ is the completely antisymmetric tensor}, $\ket{u_n(\bm{k})}$ is the Bloch wave function of quasiparticles, and $P$ is the set for occupied bands:
\begin{gather}
P\equiv\Set{n|E_n(\bm{k})<0}.
\end{gather}
The analytic formula for the Chern number has been obtained in \Ref{Daido2016}:
\begin{gather}
\nu=\sum_{(\pm,\,\bm{k}_0)}\frac{1}{2}\sgn\left[\frac{\partial(\psi\pm\bm{d}\cdot\hat{g})/\partial k_\parallel}{\muB\bm{H}\cdot\hat{g}\times\bm{d}/\alpha}\right]_{\bm{k}=\bm{k}_0},\label{TSCchern}\\
E_\pm(\bm{k}_0)=\psi(\bm{k}_0)\pm\bm{d}(\bm{k}_0)\cdot\hat{g}(\bm{k}_0)=0,\label{nodalcondition}
\end{gather}
under the conditions given by \Eqss{lowmagcond1}-\eqref{lowmagcond3}. 
The system under consideration can be regarded as a set of massive Dirac cones as illustrated in \Fig{bulkband}(b).
Equation \eqref{TSCchern} shows that the Chern number is represented by Berry curvature around the massive Dirac cones 
placed at originally nodal points given by \Eq{nodalcondition}. 
The formula (\ref{TSCchern}) is reasonable because the Chern number is ensured to be zero by the chiral symmetry in the absence of the Zeeman field and the low Zeeman field changes the Berry curvature only around the Dirac points\cite{Daido2016}. 

 \begin{figure}[htbp]
  \centering
  \begin{tabular}{ll}
      (a) $\theta=0$ & (b) $\theta=0$\\
    \includegraphics[width=40mm]{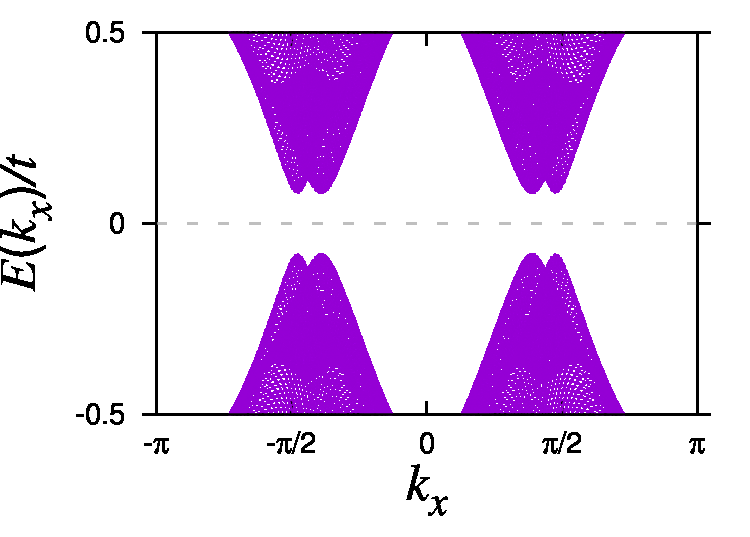} &
    \includegraphics[width=40mm]{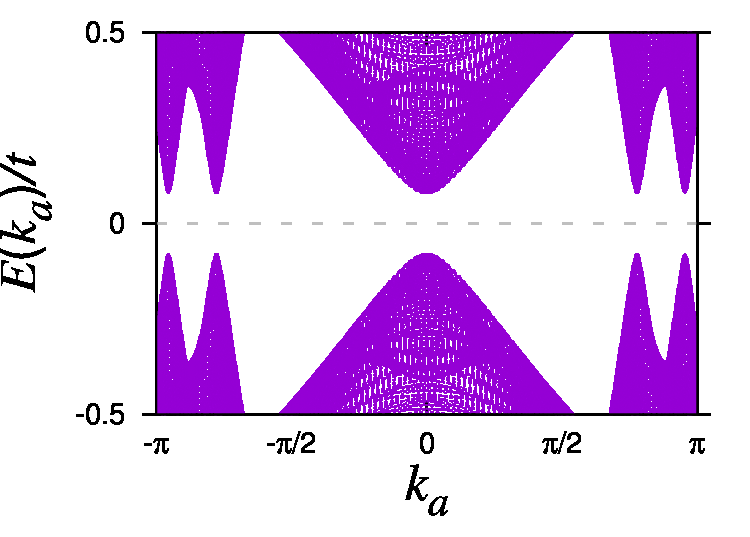}\\
      (c) $\theta=0$ & (d) $\theta=0$\\
      \includegraphics[width=40mm]{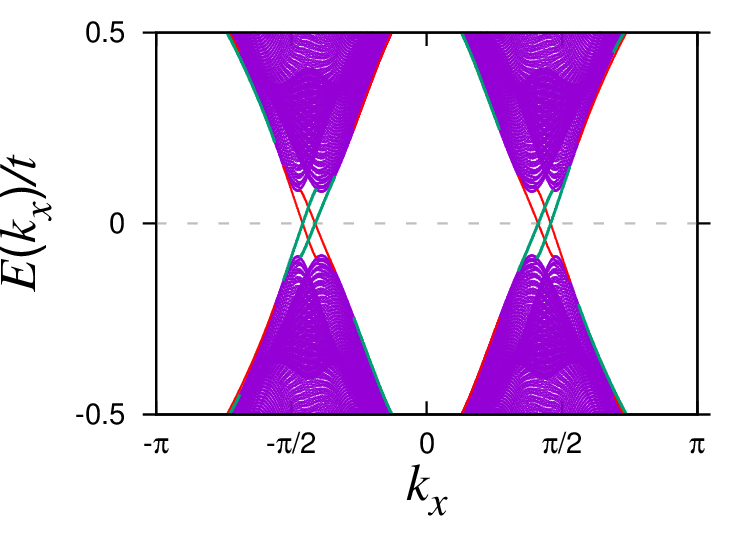}&
      \includegraphics[width=40mm]{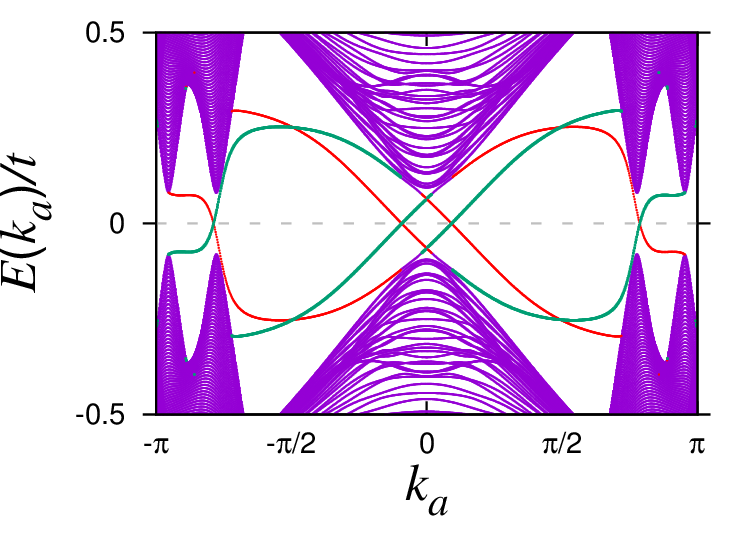}
    \end{tabular}
    \caption{
      (Color online) Energy spectrum of a $D+p$-wave SC under the perpendicular Zeeman field. Parameters used are the same as \Fig{bulkband}(b). 
(a,b) Bulk energy spectrum projected onto the edge BZ for (a) the $(010)$ edge and (b) the $(1\bar{1}0)$ edge. 
(c,d) Spectrum of ribbon-shaped systems for open boundary condition with (c) the $(010)$ edge and (d) the $(1\bar{1}0)$ edge.
Red lines show edge modes on one of the edges, and green ones show those on the opposite side. More precisely, red points indicate the states with mean position $\braket{\hat{y}}, \braket{\bm{b}} < L/4$ ($L=200$ is the number of lattice sites), while green points indicate the satates with mean position $\braket{\hat{y}}, \braket{\bm{b}} > 3L/4$. Gray dashed lines show $\mathcal{E}=0$.}
  \label{chiraledge}
\end{figure}

\begin{figure}
  \centering
  \includegraphics[height=33mm]{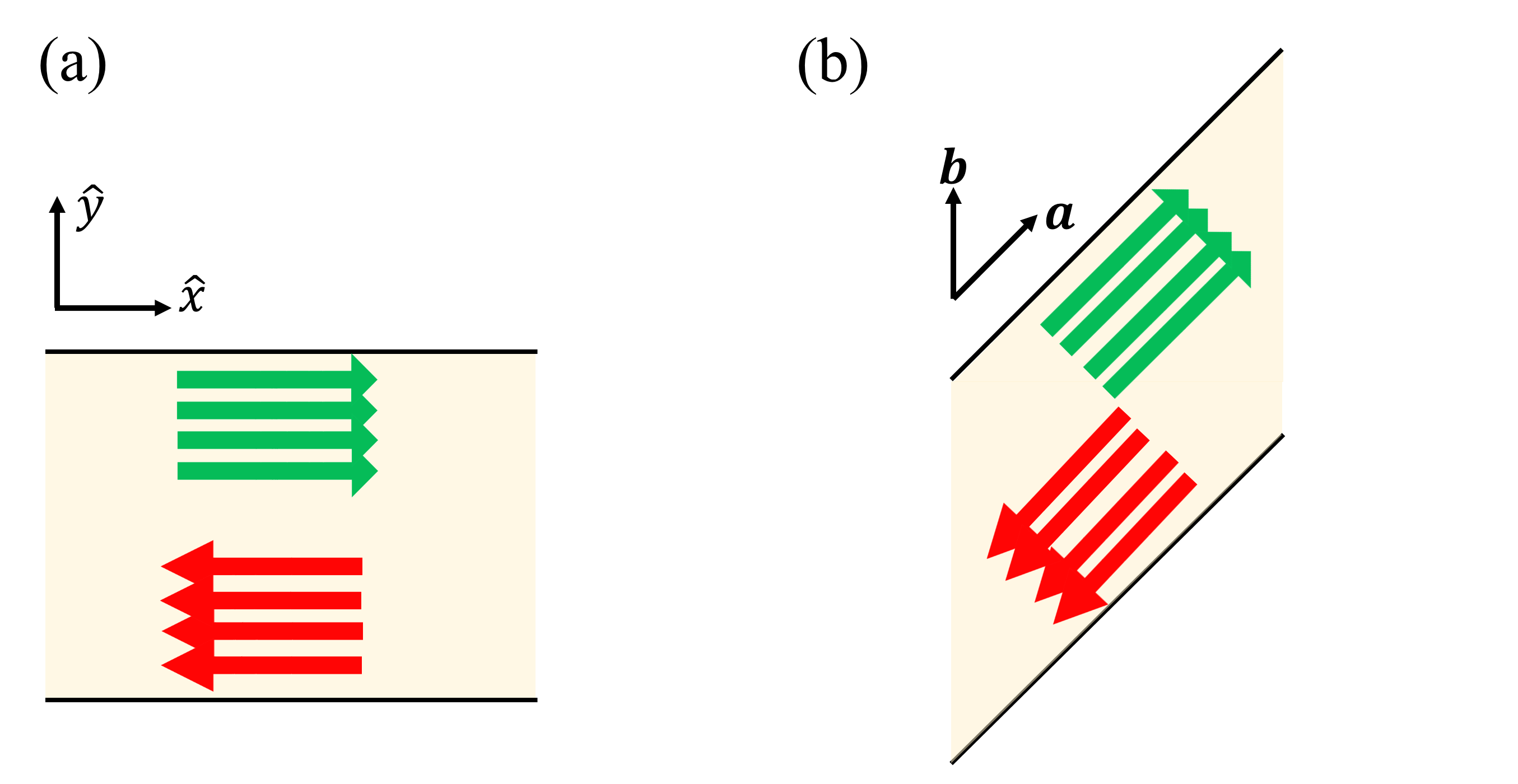}
  \caption{(Color Online) Illustration of edge states on (a) the $(010)$ edge and (b) the $(1\bar{1}0)$ edge.
The arrows represent the edge modes shown by the same color in \Fig{chiraledge}. 
}
  \label{crystalchiral}
\end{figure}
 
In the previous study we have shown that the Chern number is finite in most of the spin-singlet-dominant SCs.\cite{Daido2016} 
Then, the chiral Majorana edge states appear regardless of the boundary direction. For instance, our numerical result for the $D+p$-wave superconductivity shows chiral edge modes on both $(010)$ and $(1\bar{1}0)$ edges: \Figss{chiraledge}(a) and \ref{chiraledge}(b) show the bulk energy spectrum projected onto the $k_x$ and $k_a$ axis, respectively, while \Figss{chiraledge}(c) and \ref{chiraledge}(d) show the energy spectrum of the ribbon-shaped system with open boundary condition in the $\hat{y}$ and $\bm{b}$ direction, respectively. The configuration of edge states in real space is schematically shown in \Figss{crystalchiral}(a) and \ref{crystalchiral}(b).

As shown by our previous study, the Chern number is finite in the wide parameter regime for spin-singlet-dominant SCs\cite{Daido2016}, and therefore, the gapful topological superconductivity accompanied by the chiral Majorana edge states is one of the ubiquitous topological phenomena in originally nodal SCs gapped by the Zeeman field.

\subsection{Unidirectional Majorana edge states in gapless topological superconducting phase}
\label{subsec:UME}
When the Zeeman field is tilted from the perpendicular direction illustrated in \Fig{angle}(a), 
the condition given by \Eq{gapfulcond} will be broken. Then, the energy spectrum becomes gapless and the Chern number is ill defined.
Nevertheless, we can define another topological number $\tilde{\nu}$ by a straightforward extension of $\nu$:
\begin{gather}
\tilde{\nu}\equiv\left.\nu\right|_{P\to\widetilde{P}},\\
\widetilde{P}\equiv\Set{n|\,\text{hole bands}}.\label{conditiontilde}
\end{gather}
{Here, hole bands mean the lower $N$ bands, provided the BdG Hamiltonian is a $2N\times 2N$ matrix. In the present case, $N=2$.}
We call $\tilde{\nu}$ the band Chern number. {In \Eq{conditiontilde}, ``hole bands'' are well defined} because the band gap between the electron bands and the hole bands is given by the finite mass term $2|\muB\bm{H}\cdot\hat{g}\times\bm{d}/\alpha g|$, even when the bands cross the Fermi level [see \Fig{fig0}(d)]. 
The formula of the Chern number for gapful states is straightforwardly extended to the band Chern number in the gapless states, and indeed \Eq{TSCchern} gives the band Chern number. This is simply proved by repeating the derivation of \Eq{TSCchern}.  

\begin{figure}
  \centering
\includegraphics[width=80mm]{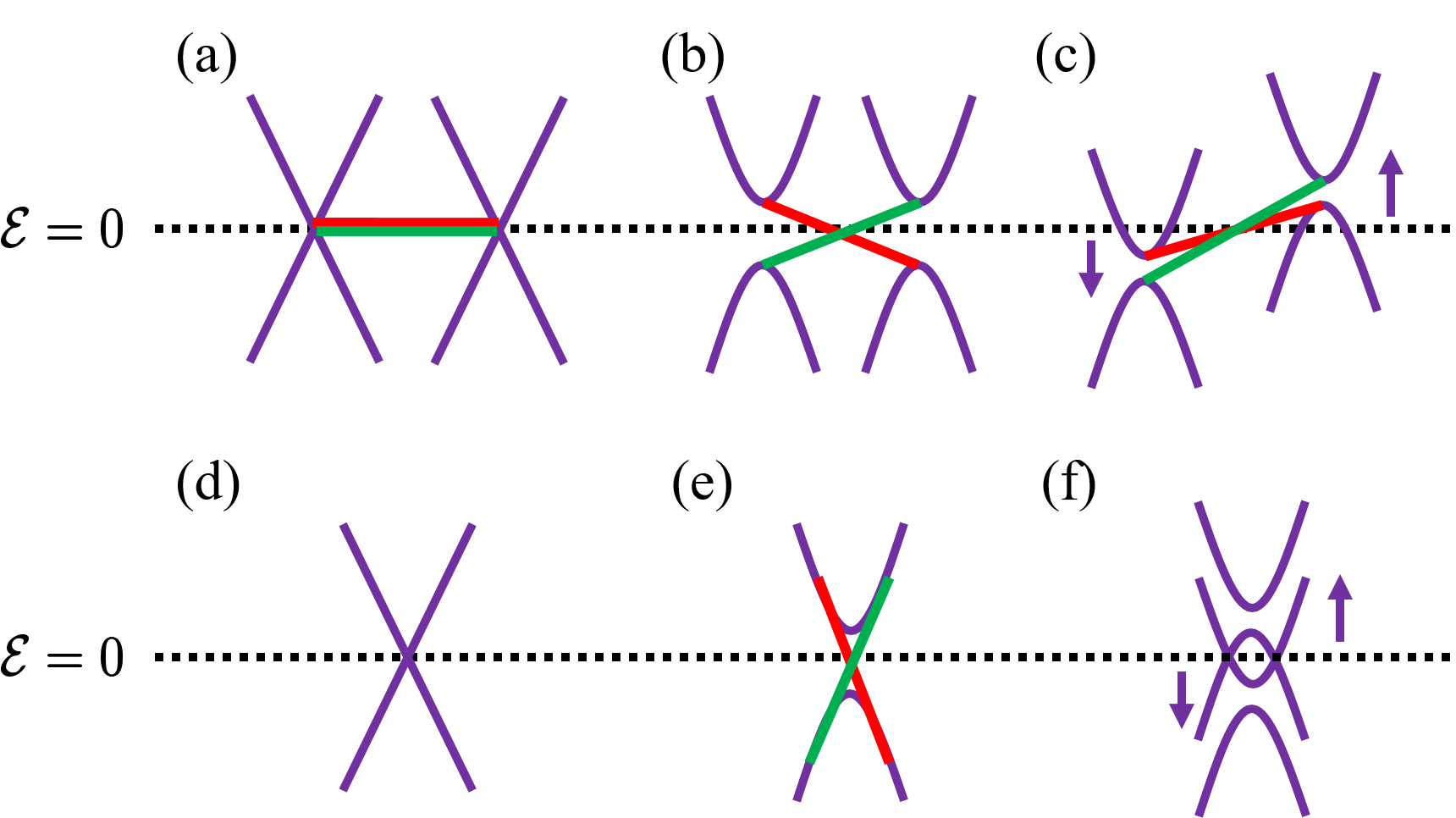}
\caption{Schematic figures for unidirectional Majorana edge states.
  Bulk bands are shown by purple, edge states by red, and those on the oppositely-oriented edge by green.
(a) Majorana flat bands, (b) chiral Majorana edge states in a gapful TSC, and (c) their residue in a gapless phase, for boundaries hosting Majorana flat bands at $H=0$. (d)-(f) Those on the boundary which does not host Majorana flat bands. 
(f) Edge states are absorbed into the bulk states in the gapless phase.}
  \label{residual}
\end{figure}

The band Chern number characterizes {\it unidirectional Majorana edge states} in the gapless phase.
They are unidirectional in the sense that edge states in oppositely oriented edges have the same chirality.
Indeed, we can see that unidirectional Majorana edge states naturally appear under tilted Zeeman field.
Let us consider a boundary which hosts Majorana flat bands in the absence of Zeeman field [\Fig{residual}(a)]. 
Then, the flat bands connect projections of nodal points. 
Under the perpendicular Zeeman field, the chiral Majorana edge states naturally come out between two massive Dirac cones [\Fig{residual}(b)]. 
When the bulk spectrum is shifted to become gapless, the edge states may remain to exist, as shown in \Fig{residual}(c), because the band touching {\it in the bulk} is prohibited.
Thus, unidirectional Majorana edge states naturally appear as a residue of chiral Majorana edge states.
{Although we have considered tilting from the perpendicular direction, the above discussions are valid even for a nearly parallel field, since the band Chern number remains well defined and finite.}

\begin{figure*}[htbp]
  \centering
    \begin{tabular}{llll}
      (a) $\theta=\pi/4$, $\phi=3\pi/4$ & (b) $\theta=\pi/4$, $\phi=3\pi/4$ & (e)&(f) $\theta=\pi/4$, $\phi=0$\\
      \includegraphics[width=40mm]{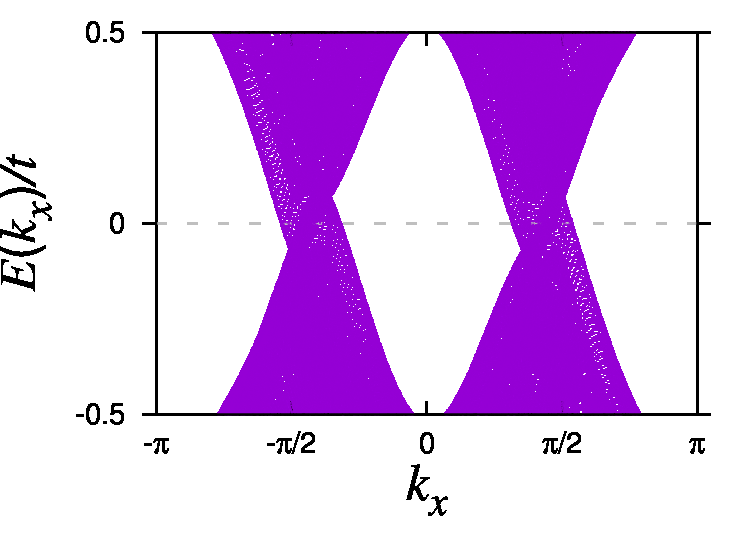} &
      \includegraphics[width=40mm]{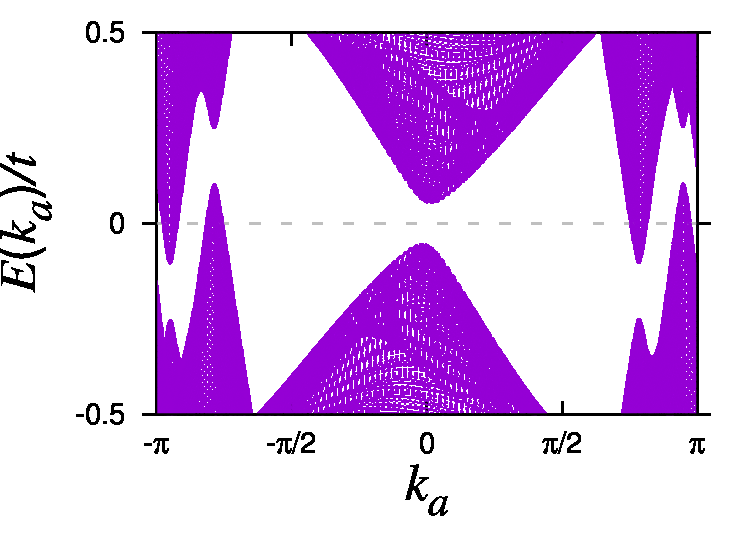} &
      \includegraphics[width=40mm]{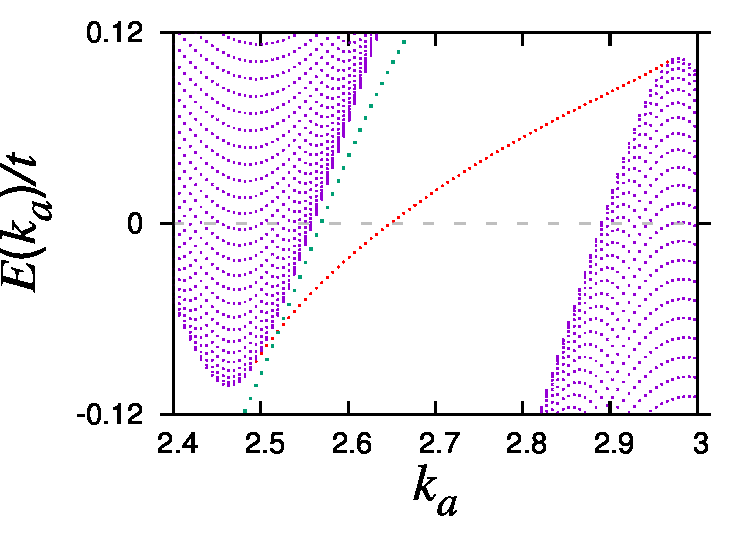} &
      \includegraphics[width=40mm]{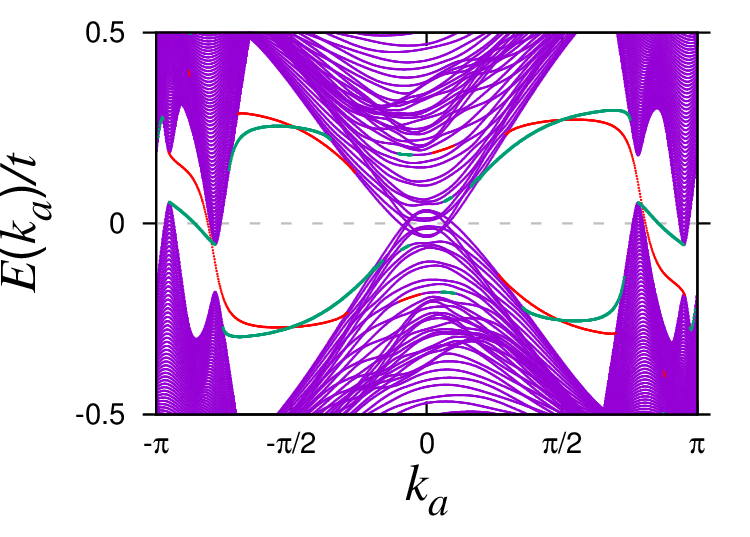}\\
      (c) $\theta=\pi/4$, $\phi=3\pi/4$&(d) $\theta=\pi/4$, $\phi=3\pi/4$&(g)&\\
      \includegraphics[width=40mm]{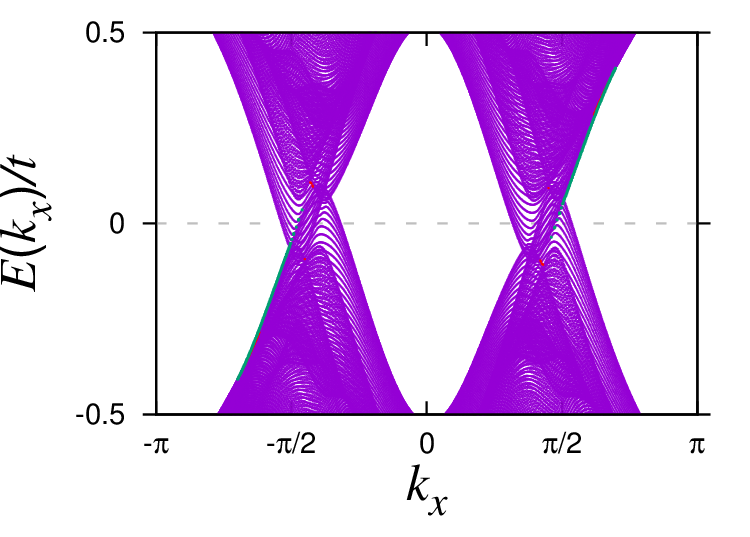}&
      \includegraphics[width=40mm]{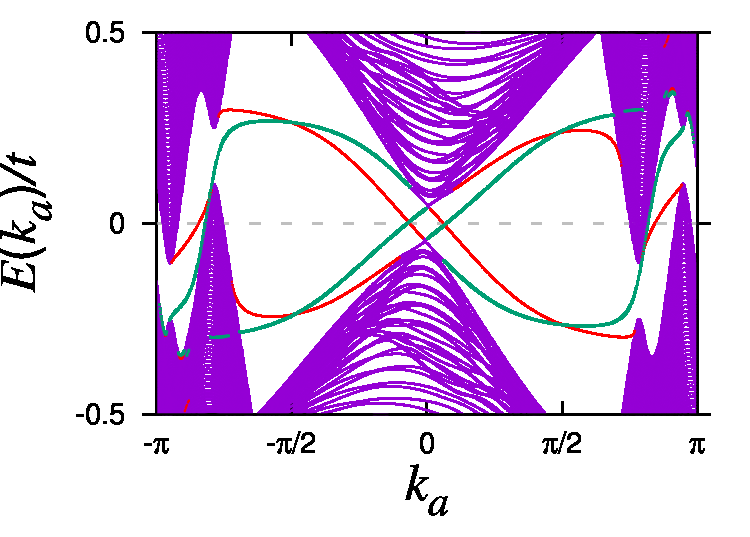}&
      \multicolumn{2}{l}{\includegraphics[height=30mm]{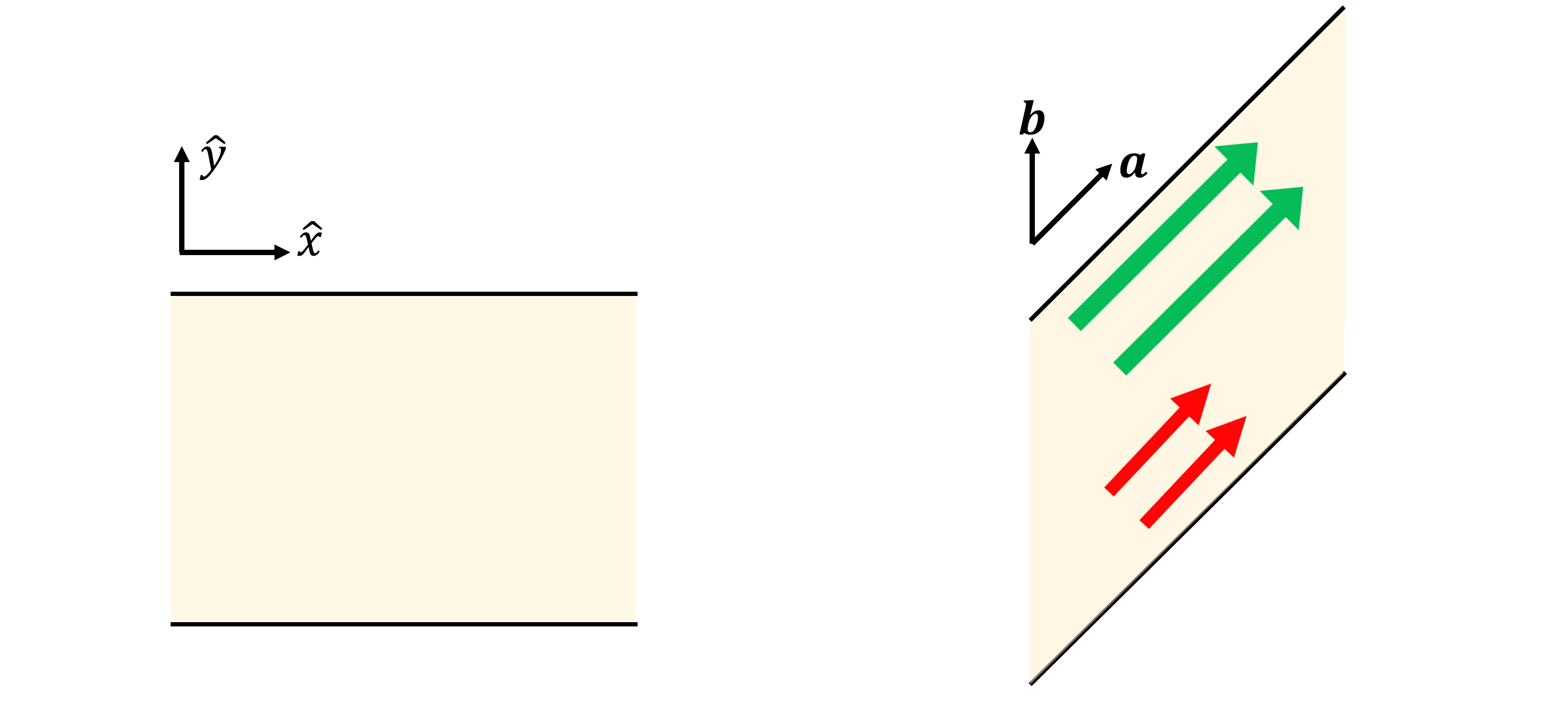}}
      \\
    \end{tabular}
    \caption{(Color online) (a)-(f) Energy spectrum of a $D+p$-wave SC in the gapless phase. Parameters are the same as \Fig{bulkband}(c). 
(a),(b) Bulk energy spectrum projected onto the edge BZ for (a) the $(010)$ edge and (b) the $(1\bar{1}0)$ edge. 
(c),(d) Spectrum of ribbon-shaped systems for open boundary condition along (c) the $(010)$ edge and (d) the $(1\bar{1}0)$ edge. 
The meaning of red and green lines is the same as \Fig{chiraledge}. 
(e) Enlargement of (d) around $k_a\sim\pi$. 
(f) Energy spectrum for the azimuth angle of the Zeeman field $\phi=0$. The others are same as (d).
(g) Illustration of edge states in (c) (left panel) and (d) (right panel) in real space. In the right panel, only unidirectional edge states around $k_a\sim\pm\pi$ are shown, for simplicity.}
\label{asymmetricMajorana}
\end{figure*}
An important difference from the chiral Majorana edge states is that the existence of unidirectional Majorana edge states 
depends on the boundary direction. The boundary hosting the unidirectional Majorana edge states is almost the same as 
that for the Majorana flat bands.
{For example, let us consider a boundary direction hosting a Dirac spectrum around which several Dirac cones in the bulk are overlapped and the 1D winding number given by \Eq{1DwindGamma} vanishes.}
This is the case of a $D+p$-wave SC with a $(010)$ edge. 
In such case, the Majorana flat bands do not appear at $H=0$ [\Fig{residual}(d)]. 
In the gapful topological superconducting state, the chiral edge states would be like \Fig{residual}(e). 
When the system becomes gapless, band overlapping in the surface BZ would occur, as shown in \Fig{residual}(f), 
and then the edge states no longer exist. 
In this way, the unidirectional Majorana edge states can also be regarded as a residue of the Majorana flat bands. 
When the latter appears on a boundary, the gapless topological phase is accompanied by the former. 


The nontrivial band Chern number does not prohibit {\it band overlapping} in the {\it surface BZ}, although it prohibits {\it band touching} in the {\it bulk}. Even if bulk bands are well-separated in the 2D momentum space, the projected bands onto a $k$ axis may overlap. In such band-overlapping situations, the residue of chiral Majorana edge states is absorbed into the bulk states. Thus, unidirectional Majorana edge states appear on the boundary avoiding band overlapping.


The transition between the Majorana flat bands, the chiral Majorana edge states, and the unidirectional edge states presented {above} is quite general for noncentrosymmetric spin-singlet SCs with a nodal gap. {Actually, the above discussion assumes only the presence of Majorana flat bands in the absence of the Zeeman field, and the Majorana flat bands generally exist in nodal noncentrosymmetric SCs when we choose appropriate boundary directions, as shown in \Sec{sec:ubiquitous}.}
{Next, we illustrate the appearance of} unidirectional Majorana edge states in the model for $D+p$-wave superconductivity, given by \Eqss{D+pmodel1}-\eqref{D+pmodel2}.
The transition from the Majorana flat bands and chiral Majorana edge states to the unidirectional edge states is clarified.

First, we show the bulk energy spectrum in \Figss{asymmetricMajorana}(a) and \ref{asymmetricMajorana}(b), which are obtained by the projection 
onto the $\hat{x}$ and $\bm{\alpha}$ directions, respectively. 
The tilting angle of the Zeeman field, $\theta =\pi/4$ and $\phi=3\pi/4$, leads to the gapless topological 
superconducting state with the nontrivial band Chern number $\tilde{\nu}=-4$. 
It should be noticed that the band overlapping occurs on the $(010)$ edge, but not on the $(1\bar{1}0)$ edge.

Figures~\ref{asymmetricMajorana}(c) and \ref{asymmetricMajorana}(d) show the spectrum of ribbon-shaped systems with open boundary condition 
corresponding to \Figss{asymmetricMajorana}(a) and \ref{asymmetricMajorana}(b), respectively. 
We indeed see the unidirectional Majorana edge states only on the $(1\bar{1}0)$ edge [\Fig{asymmetricMajorana}(d)]. 
In this case, smooth transitions in the edge states are clearly observed. 
Majorana flat bands in \Fig{flatbands}(d) are deformed into the chiral Majorana edge states in \Fig{chiraledge}(d) 
and unidirectional Majorana edge states in \Fig{asymmetricMajorana}(d) under the Zeeman field. 
On the other hand, the chiral edge states on the $(100)$ edge [\Fig{chiraledge}(c)] are absorbed into the overlapped 
bulk bands [\Fig{asymmetricMajorana}(c)]. 

Although edge states around $k_a\sim0$ are counterpropagating and similar to the chiral Majorana edge states in \Fig{chiraledge}(d), those around $k_a=\pm\pi$ are remarkably altered from \Fig{chiraledge}(d): 
A chirality of the edge mode changes, and the edge states propagate in the same direction on both sides of the edges, as schematically depicted in \Fig{asymmetricMajorana}(g). 
Thus, \Fig{asymmetricMajorana}(d) shows an example of unidirectional Majorana edge states: the net chirality of the two edge modes is unidirectional, although the other two modes are counterpropagating.

It should be noted that the edge current of unidirectional edge states is not conserved within themselves.  
However, the gapless bulk states compensate the edge current and the total current may be conserved, as pointed out in \Ref{Wong2013}.
For this reason, unidirectional Majorana edge states are unique to gapless topological superconducting phases.

So far, we have analyzed properties of unidirectional Majorana edge states for the azimuths angle of the Zeeman field 
$\phi=3\pi/4$. In general, edge states depend on $\phi$. 
Figure~\ref{asymmetricMajorana}(f) shows the energy spectrum on the $(1\bar{1}0)$ edge for $\phi=0$. 
In this case, edge states around $k_a\sim0$ are absorbed into the bulk states because of the band overlapping. 
On the other hand, the unidirectional edge states around $k_a\sim\pm\pi$ are robust.
As illustrated above, a part of the edge states may disappear owing to the band overlapping.
Even then, the other part of the edge states is separated from bulk bands when a suitable field direction is chosen.

Related to the topological phase transition, we comment on the case of the parallel Zeeman field $\theta=\pi/2$ 
illustrated in \Fig{angle}(c). Then, the mass term giving rise to the gap in the Dirac spectrum vanishes. 
Accordingly, the band Chern number changes from $-4$ to $4$ through $\theta=\pi/2$. 
It is proven that the band Chern number must change the sign at $\theta=\pi/2$, 
simply because the field $H$ with $\pi/2<\theta<\pi$ is equivalent to the field $-H$ with $0<\theta<\pi/2$. 
Thus, $\theta=\pi/2$ is a critical point of the topological phase transition associated with a change of 
the topological invariant. 

{Finally, we comment on previous studies of unidirectional edge states\cite{Wong2013,Baum2015}. 
The first work by Wong {\it et al}.\cite{Wong2013} considers $s+p$-wave superconductivity, and the unidirectional Majorana edge states appear in an extraordinarily high Zeeman field $\muB H\sim|\mu|$. 
Their model is characterized by the band Chern number $\pm1$
, although the band gap is closed owing to the specific choice of the $d$ vector.
In another work by Baum {\it et al}.\cite{Baum2015}, $s$-wave superconductivity with coexisting spatially modulated magnetic order has been studied, indicating unidirectional Majorana edge states for some edge termination and disorder strength. They proposed the surface of 3D topological insulators with spiral magnetic order as a potential platform, but experimental realization by magnetic doping remains to be a future issue.
In contrast to these proposals, our model covers almost all the 2D nodal noncentrosymmetric spin-singlet SCs 
under low Zeeman fields, and may be more easily set up in experiments. Therefore, our proposal would be more suitable for experimental realization of unidirectional Majorana edge states.}

\section{Ubiquitousness of unidirectional Majorana edge states in noncentrosymmetric systems}
\label{sec:ubiquitous}
In this section, we clarify the generality of our scenario for unidirectional Majorana edge states.
In the previous section, the appearance of unidirectional Majorana edge states has been discussed based on the existence of Majorana flat bands in the absence of the Zeeman field.
The nodal $d$-wave superconductivity has been dealt with as an example.
However, the existence of Majorana flat bands and unidirectional Majorana edge states may still be unclear for other pairing symmetry.
In the following, we prove that we can always find boundary directions hosting Majorana flat bands in noncentrosymmetric nodal superconductivity, irrespective of the pairing symmetry.
Then, the discussion parallel to \Sec{subsec:UME} brings about the conclusion that unidirectional Majorana edge states are ubiquitous in noncentrosymmetric nodal SCs under Zeeman field.

First, we discuss boundary directions hosting Majorana flat bands.
The key idea for choosing the boundary is quite simple: we can always choose an axis $k_a'$ on the reciprocal lattice space so that all the point nodes in the first Brillouin zone are projected onto different points on the $k'_a$ axis.
It has been shown that each of the nodes is protected by the winding number given by \Eq{formulawind2}, taking $\pm1$, unless nodes are accidentally placed on TRIM. Each projection of nodes onto the $k_a'$ axis accompanies the change of the winding number $W(k_a')$ by $\pm1$. Thus, a nontrivial value of $W(k_a')$ in some region is ensured. It means the existence of Majorana flat bands on the edge. In the following, we call such boundary hosting Majorana flat bands as the flat-band boundary (FBB).

For example, one of the FBBs in $D+p$-wave superconductivity is the $(120)$ edge. Corresponding reciprocal lattice vectors are given by $\bm{\alpha}'=\hat{x},\quad \bm{\beta}'=-2\hat{x}+\hat{y}$.
We can use \Fig{exwind} to check that all the projections of nodes are not overlapped on the $k_a'$ axis.
In the same way, we can show that all the noncentrosymmetric nodal SCs, such as extended $S+p$-wave superconductivity or $s+P$-wave superconductivity, host Majorana flat bands on FBBs.

The connection of Majorana flat bands and chiral Majorana edge states in \Fig{residual} is also irrespective of pairing symmetry. Taking FBBs, we can interpret chiral Majorana edge states under the perpendicular Zeeman field as smoothly-modified Majorana flat bands.

An advantage of considering FBBs is that all the Dirac cones are not overlapped in the surface BZ.
This makes a sharp contrast to the $(1\bar{1}0)$ surface in the $D+p$-wave superconductivity, where four point nodes in the bulk are overlapped at $k_a=0$.
At the bulk nodal point $\bm{k}_0$, the mass term and paramagnetic shift of the Dirac spectrum are governed by a single vector $\hat{g}(\bm{k}_0)$. Thus, band-overlapping leading to the disappearance of edge states such as \Fig{residual}(f) does not occur, in contrast to \Fig{asymmetricMajorana}(f).
From this fact, we understand a simple bulk-edge correspondence:
{\it Finite band Chern number ensures the existence of edge states on FBBs.}
Now, let us consider two massive Dirac cones connected by chiral Majorana edge states like in \Fig{residual}(b). Assuming electronic spins are polarized parallel to $\hat{g}(\bm{k}_1)$ in the left cone, while $\hat{g}(\bm{k}_2)$ in the right cone, we can realize the unidirectional Majorana edge states like in \Fig{residual}(c), by applying the Zeeman field satisfying
\begin{gather}
  \mu_B\bm{H}\cdot(\hat{g}(\bm{k}_1)+\hat{g}(\bm{k}_2))=0.
\end{gather}
Then, the paramagnetic shift of the two Dirac cones is in the opposite direction. Thus, unidirectional Majorana edge states generally appear on FBBs.

An exceptional case is the $D_{3h}$ and $C_{3h}$ point group symmetry, where the $g$ vector is parallel to the $z$ axis in the whole Brillouin zone. Then, an additional condition, $\hat{g}(\bm{k}_1)=-\hat{g}(\bm{k}_2)$, has to be satisfied for the opposite paramagnetic shift.

Thus, Majorana flat bands, chiral Majorana edge states, unidirectional Majorana edge states, and topological phase transitions between these phases are ubiquitous phenomena in noncentrosymmetric nodal SCs.

\section{Conclusions and discussions}
\label{sec:conclusion}
We summarize the results obtained in this paper.
We have studied noncentrosymmetric time-reversal symmetric 2D nodal SCs and have examined the topological superconductivity caused by the broken time-reversal symmetry due to a low Zeeman field. The gapful-gapless transition in the bulk spectrum against tilting of the field has been elucidated. The gapful phase is a strong topological superconductor of class D in the large parameter range. 
The critical Zeeman-field angle is estimated to be about $\theta = 6^\circ$ for high-$T_c$ cuprate SCs. This result reveals the stability of the paramagnetically induced gapful topological superconductivity reported previously\cite{Yoshida2016,Daido2016} and points to the experimental realization.

We clarified the topological edge states in the gapful and gapless states. 
In the absence of the Zeeman field, Majorana flat bands may appear on a boundary of certain direction. We proposed a compact formula for calculating the winding number protecting the flat bands. 
In the perpendicular Zeeman field, chiral Majorana edge states appear in the gap of the bulk spectrum, regardless of the boundary direction.
In the gapless phase under the tilted Zeeman field, the Chern number is ill-defined. However, the well-defined band Chern number characterizes the topologically nontrivial properties. 
The nontrivial band Chern number specifies unidirectional Majorana edge states on the boundary avoiding the overlapping of bulk bands.
The condition for such boundary directions has been discussed.
Unidirectional Majorana edge states propagate in the same direction on both sides of the edges. 
These topological phases and topological edge states have been demonstrated by analyzing the model for $D+p$-wave SCs.

It is stressed that the results obtained in this paper are applicable to almost all of the spin-singlet-dominant noncentrosymmetric 2D nodal SCs.
In particular, heterostructures of high-$T_{\rm c}$ cuprate SCs and ferromagnets\cite{Chakhalian2006,Satapathy2012,Uribe-Laverde2014,Sen2016,Das2014} are platforms realizing the topological superconductivity. 
Thin films of cuprate SCs\cite{Bollinger2011,Garcia-Barriocanal2013,Werner2010,Jin2016,Leng2011,Zeng2015,Nojima2011} and heavy-fermion SCs\cite{Izaki2007,Shimozawa2014} under external magnetic fields may also be promising candidates, because the paramagnetic effect is dominant as ensured by the large Maki parameter.
These materials are known to host $d$-wave superconductivity in the bulk\cite{Yanase2003}.The presence and importance of inversion-symmetry breaking in superconducting heterostructures have already been evidenced by the upper critical field measurements\cite{Goh2012,Shimozawa2014}. Although admixing of spin-triplet order parameter has not been directly observed, inversion-symmetry breaking immediately means parity admixing of the order parameter from the group theoretical analysis. Furthermore, the parity admixing has been demonstrated by the analysis of microscopic models for spin-fluctuation-mediated superconductivity\cite{Bauer2012}. Here we stress that the details of the admixing order parameter are not important for the appearance of unidirectional Majorana edge states\cite{Daido2016}.

{\it Note}: Recently, we noticed a related paper \cite{Hao2016} in which the band Chern number and unidirectional edge states in topological $s$-wave superconductivity\cite{Sato2009_STF,Sato2010_STF,Sau2010,Alicea2010,Lutchyn2010,Mourik2012,Nadj-Perge2014} are studied. 
However, the nodal noncentrosymmetric SCs which ubiquitously show a variety of topologically superconducting states are studied in this paper. 
\section*{Acknowledgements} 
The authors are grateful to M.~Nakagawa, S.~Ikegaya, and A.~Shitade for fruitful discussions. 
This work was supported by ``J-Physics''(Grant No. JP15H05884) and ``Topological Materials Science'' (Grant No. JP16H00991) Grant-in Aid for Scientific Research on Innovative Areas from MEXT of Japan, and by JSPS KAKENHI Grants No. JP15K05164, and No. JP15H05745.
%

\end{document}